\documentclass[12pt,a4paper]{article}

\usepackage[utf8]{inputenc}
\usepackage[T1]{fontenc}
\usepackage{amsmath,amssymb}
\usepackage{graphicx}
\usepackage{subcaption}
\usepackage{cite}
\usepackage{hyperref}
\usepackage[margin=2.5cm]{geometry}
\usepackage{setspace}
\usepackage{siunitx}
\usepackage{booktabs}
\usepackage{tabularx}
\usepackage{multirow}
\usepackage{array}
\usepackage{algorithm}
\usepackage{algpseudocode}

\title{Physics-Informed Attention Mechanism and Generalization Capability of Deep Learning-Based Grain Growth Evolution Prediction}

\author{Pungponhavoan ~Tep\thanks{Corresponding author: pungponhavoan.tep@minesparis.psl.eu} ,
Marc ~Bernacki \\
	\\
	Mines Paris, PSL University\\
 	Centre for Material Forming (CEMEF), UMR CNRS\\
 	06904 Sophia Antipolis, France\\
}

\date{}

\begin{document}

\maketitle

\begin{abstract}
Machine Learning (ML) models for grain growth prediction are typically trained on idealized synthetic data, yet practical applications require generalization to conditions outside the training distribution. This study evaluated the Out-Of-Distribution (OOD) generalization capability of the trained model from our previous study across three test cases, including experimental microstructures, microstructures characterized by a bimodal grain size distribution, and abnormal grain growth. To further probe whether physics-informed architectural design could improve robustness under these different conditions, a boundary-masked attention mechanism was proposed specifically for grain growth, constraining attention to grain boundary pixels. Both the baseline and the proposed physics-informed attention model were evaluated without retraining or fine-tuning on the OOD data. Both models successfully generalized to all three test cases, yet the boundary-masked attention mechanism provided substantial improvements, with the most notable gains for microstructures characterized by a bimodal grain size distribution, where Structural Similarity Index Measure (SSIM) improved from \num{0.6221} to \num{0.7609} and mean grain size ($\overline{R}$) error decreased from \SI{8.75}{\percent} to \SI{3.57}{\percent}. The attention heatmap analysis revealed that the boundary-masked attention model learned to concentrate attention on large grain boundaries in a manner consistent with curvature-driven grain growth physics, emerging from training without being explicitly encoded into the architecture. These results indicate that models trained on synthetic data can generalize to diverse OOD conditions without retraining, and that physics-informed attention may improve accuracy when the boundary morphology matches the training domain.

\end{abstract}

\section{Introduction}

Data-driven approaches have emerged as computationally efficient alternatives to traditional physics-based simulation methods in the field of materials science, offering substantial reductions in computation time across a wide range of simulation tasks. Grain growth, for example, is a thermomechanical process in which the migration of grain boundaries is driven by the reduction of total boundary energy, causing larger grains to progressively consume smaller ones and resulting in an increase in mean grain size over time. Traditional approaches including phase-field~\cite{yang2021phase, moelans2008quantitative, krill2002computer, chen2002phase, steinbach2009phase}, level-set~\cite{bernacki2024kinetic, murgas2021comparative} predict the phenomenon by iteratively solving Partial Differential Equations (PDEs) to calculate grain boundary migration at each timestep. This iterative computation across every point in the computational domain at each timestep renders the mentioned methods computationally expensive and time-consuming, with computation times ranging from hours to days for small-scale simulations and extending to weeks for industrial-scale ones. Our previous study~\cite{TEP2025121486} demonstrated the potential of data-driven approaches for grain growth prediction by developing a Machine Learning (ML)-accelerated framework, reducing computation time from the order of hours to approximately \SI{10}{\second} per hour of grain growth evolution. Despite this substantial reduction in computation time, the developed framework still achieved high prediction accuracy, with Structural Similarity Index Measure (SSIM)~\cite{SSIM2004} scores of up to \SI{86}{\percent} and a mean grain size ($\overline{R}$) error as low as \SI{0.07}{\percent} across multiple domain sizes and initial configurations.

Since the framework is a data-driven approach and the model was trained exclusively on synthetic data of pure grain growth under isotropic grain boundary properties and smooth grain boundary morphology, one fundamental question arises regarding the generalization capability of the model when applied to practical conditions not represented in the training dataset. Such conditions are commonly referred to as Out-Of-Distribution (OOD) generalization. OOD refers to inputs that differ systematically from the training data distribution, whether through different statistical properties, novel feature characteristics, or underlying processes not seen during training. Generalization to OOD scenarios is a critical challenge given that neural networks tend to perform well on data that resembles their training distribution but can fail unpredictably when confronted with inputs outside this learned distribution~\cite{shen2021towards,li2024probing}. Addressing OOD generalization is therefore critical for enabling reliable deployment of such models in practical industrial settings, where the model may encounter microstructures that differ systematically from the idealized synthetic data used during training. Yet existing ML models for microstructure evolution prediction are typically trained and evaluated exclusively under in-distribution conditions, and systematic evaluation of OOD generalization capability remains largely absent from the literature.

Against this background, this study addresses two objectives. First, this study systematically evaluated the OOD generalization capability of a grain growth ML model trained solely on pure grain growth, without any retraining or fine-tuning, across three challenging OOD test cases. These were (1) experimental microstructures, (2) synthetic microstructures characterized by a bimodal initial grain size distribution representing statistical configurations outside the training distribution, and (3) abnormal grain growth with fundamentally different evolution kinetics compared to the homogeneous grain growth used during training. These test cases represent distinct types of distribution shift that are relevant to practical industrial applications, and even critical for an abnormal configuration. Second, this study proposed a boundary-masked attention mechanism, a novel physics-informed attention mechanism design specifically for grain growth, that constrains spatiotemporal attention to grain boundary networks, introducing an inductive bias aligned with curvature-driven boundary migration. This mechanism was evaluated on the same three OOD test cases alongside the baseline model to systematically identify the conditions under which physics-informed attention provided benefits versus limited improvement, and to examine the potential mechanisms underlying these differences. This analysis provides insights into the strengths and limitations of physics-informed attention for grain growth prediction under distribution shift.

\section{Methodology}
\subsection{Baseline Model}
\label{sec:baseline_model}
The prediction model that serves as a baseline in this study is the Deep Learning (DL) framework previously developed and validated in the previous study~\cite{TEP2025121486} that was based on high-fidelity data validated against experimental data for nickel-based superalloys and stainless steel. The baseline model adopted an encoder-decoder architecture that combined Convolutional Autoencoders with Convolutional Long Short-Term Memory (ConvLSTM) networks, as illustrated in Figure~\ref{fig:architecture}(a). The encoder of the Convolutional Autoencoder compressed input microstructure images into a compact latent representation, retaining key features such as grain boundary positions and local topology while discarding noise. An encoder ConvLSTM network then learned spatiotemporal patterns of grain growth evolution from these compressed features, preserving the spatial structure of the microstructure throughout the temporal sequence, while a decoder ConvLSTM network generated the predicted latent representations autoregressively. The decoder of the Convolutional Autoencoder subsequently reconstructed these predicted representations back to full-resolution microstructure images of grain growth evolutions.

In the previous study, the framework was trained on three model configurations with different temporal window sizes, namely S-10-10, S-20-20, and S-30-30, where the numbers indicate the input and output sequence lengths in minutes. For example, the S-30-30 configuration inputs \num{30} consecutive frames at one-minute intervals and predicts the subsequent \num{30} frames of evolution. All three configurations were trained on an identical dataset of \num{647} synthetic pure grain growth evolutions generated from ToRealMotion (TRM)~\cite{florez2020novel,florez2021parallelization} simulations and validated against pre-existing experimental EBSD data from annealing experiments. These simulations were conducted under isotropic boundary conditions, capillarity-driven grain boundary migration, and unimodal Gaussian initial grain size distributions. This framework thus represents a precise and validated physical playground for grain growth at high temperature and at the polycrystal scale. The generated data spanned domain sizes ranging from \SI{2}{\milli\meter} $\times$ \SI{2}{\milli\meter} to \SI{5}{\milli\meter} $\times$ \SI{5}{\milli\meter} and initial grain configurations with varying grain size distributions and topological characteristics. Initial microstructure states were generated using LavoGen through Laguerre-Voronoï tessellation, with the generation parameters listed in Table~\ref{tab:dataset_gen_parameters_lavogen}.

\begin{table}[htbp]
\centering
\footnotesize
\renewcommand{\arraystretch}{1.2} 
\setlength{\extrarowheight}{2pt} 
\begin{tabularx}{\textwidth}{p{2.2cm}>{\centering\arraybackslash}p{2.5cm}X}
\toprule
\textbf{Parameter} & \textbf{Value(s)} & \textbf{Description} \\
\midrule
$a$ (\si{\milli\meter}) & \num{2}, \num{3}, \num{4}, \num{5} & Side length of the square domain. \\ \cline{1-3}
Distribution & $\mathcal{N}(\overline{R},\,\sigma^{2})$ & The type of distribution used for the ECR distribution (Laguerre-Voronoï tessellation \cite{BERNACKI2024101224,Hitti2012}) representative of experimental data. \\ \cline{1-3}
$\overline{R}$ $(\SI{}{\micro\meter})$ & \num{20} & The mean value of the ECR distribution. \\ \cline{1-3}
$\sigma$ $(\SI{}{\micro\meter})$ & \num{2}, \num{4}, \num{8}, \num{16}, \num{32} & The standard deviation of the ECR distribution. \\
\bottomrule
\end{tabularx}
\caption{Parameters for generating initial microstructure states using LavoGen. Equivalent Circle Radius (ECR) = the radius of the circle having the same area as the considered grain (in \SI{}{\micro\meter}).}
\label{tab:dataset_gen_parameters_lavogen}
\end{table}

For this study, the previously trained model with the S-10-10 configuration was selected as the baseline model. This choice was motivated by practical considerations, as the S-10-10 model requires shorter input sequences (\SI{10}{\minute} rather than \SI{20}{\minute} or \SI{30}{\minute}), which could be more readily available in practical applications. Collecting long temporal sequences presents two distinct constraints. For experimental characterization, material availability and characterization logistics limit the number of consecutive frames that can be acquired, while for PDE-based simulation data, generating longer sequences increases computational cost. Despite requiring shorter input window sizes, the S-10-10 model maintained prediction accuracy comparable to models trained with longer temporal windows, demonstrating that it effectively captured grain growth dynamics even from limited temporal context. It should be emphasized that the baseline model weights used in this study are the original trained weights from the previous study, loaded directly without modification. No retraining, fine-tuning, or transfer learning was performed on the selected baseline model. This allows direct evaluation of the true generalization capability of models trained exclusively on synthetic, in-distribution data when presented with fundamentally different test conditions.

\subsection{Integration of Attention Mechanisms}
\label{sec:attention_mechanisms}

The attention mechanism~\cite{Vaswani2017} is a foundational component of modern large language models and sequence-to-sequence architectures in natural language processing. The mechanism operates by computing relevance scores between a query and a set of key-value pairs, enabling a model to selectively weight the most informative parts of the input when generating each output. The scaled dot-product attention formulation~\cite{Vaswani2017} computes these scores as follows:

\begin{equation}
\text{Attention}(\mathbf{Q}, \mathbf{K}, \mathbf{V}) = \text{softmax}\left(\frac{\mathbf{Q}\mathbf{K}^T}{\sqrt{d_k}}\right)\mathbf{V},
\label{eq:attention}
\end{equation}

where $\mathbf{Q}$ (query), $\mathbf{K}$ (key), and $\mathbf{V}$ (value) represent learned projections of the input features, and $d_k$ denotes the dimensionality of the key vectors. This formulation, originating from natural language processing applications, allows models to selectively attend to different parts of an input sequence. For example, in a machine translation task, when generating the word ``chat'' in French from the English sentence ``The cat sat on the mat'', the attention mechanism learns to assign high weights to the word ``cat'' while suppressing irrelevant words such as ``mat'' or ``on''. This selective focusing enables the decoder to access contextually relevant information regardless of positional distance in the sequence.

While attention mechanisms have achieved considerable success in sequence-to-sequence tasks involving discrete tokens~\cite{Bahdanau2015}, their application to spatiotemporal microstructure prediction presents distinct challenges and opportunities. Unlike natural language processing, where attention operates over one-dimensional sequences of text, grain growth prediction requires attention over both spatial dimensions of the microstructure images and the temporal dimension of the evolution sequence. Furthermore, the physical nature of grain boundary migration suggests that not all regions of the microstructure are equally informative, as the dynamics are governed primarily by grain boundaries rather than grains.

\begin{figure}[h!]
    \centering
    \begin{subfigure}[t]{\linewidth}
    \centering
    \includegraphics[width=\linewidth]{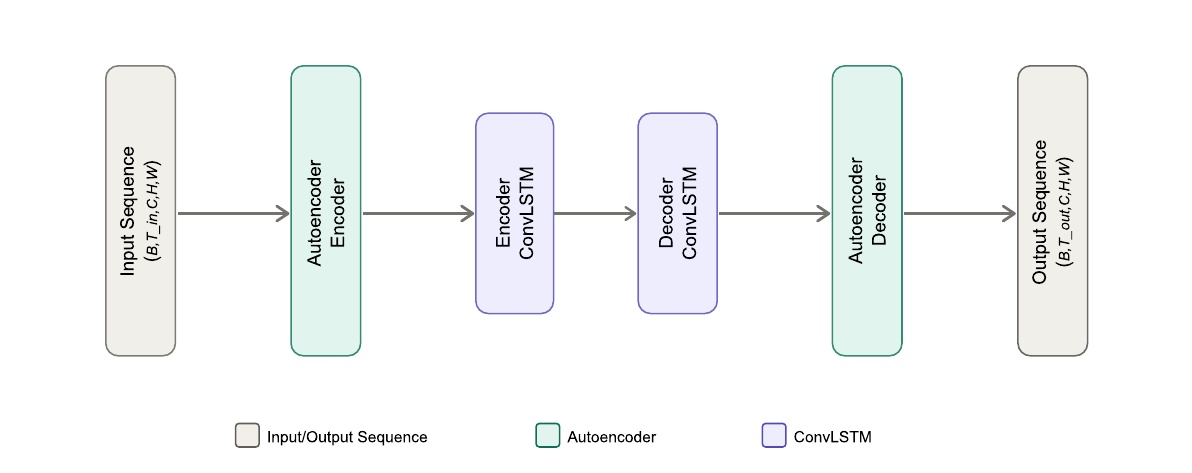}
    \caption{}
    \label{fig:architecture_baseline}
    \end{subfigure}
    \vspace{0.10cm}
    \begin{subfigure}[t]{\linewidth}
    \centering
    \includegraphics[width=\linewidth]{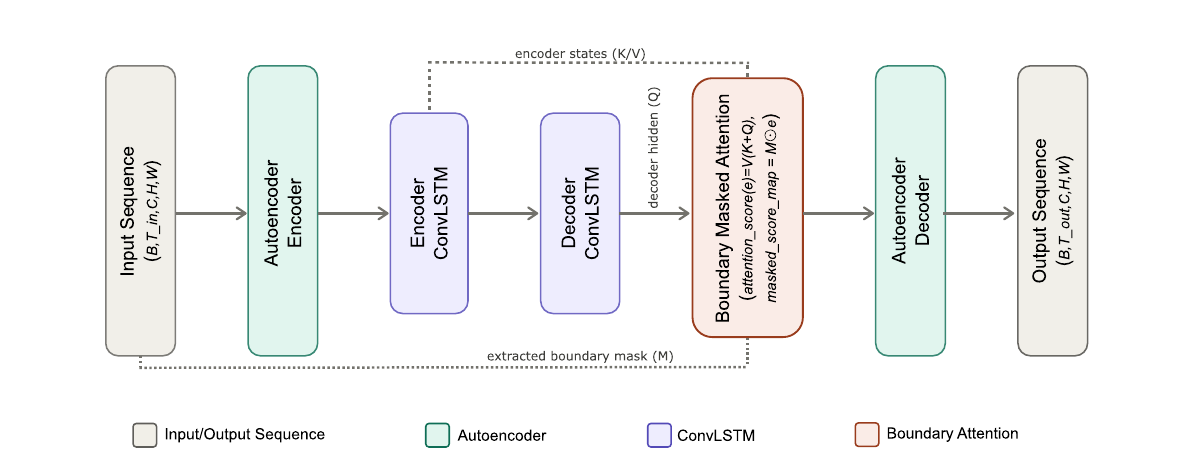}
    \caption{}
    \label{fig:architecture_attention}
    \end{subfigure}
    \caption{Neural network architectures for grain growth evolution prediction: (a) baseline Autoencoder-ConvLSTM from the previous study~\cite{TEP2025121486}; (b) boundary-masked attention model extending the baseline.}
    \label{fig:architecture}
\end{figure}

To address these considerations, a boundary-masked attention mechanism was proposed specifically for grain growth evolution prediction. The proposed mechanism computes a single normalized distribution over all $T \times H' \times W'$ positions in the encoded feature space, where $T$ is the number of input timesteps and $H' \times W'$ is the spatial resolution of the latent representation. This allows the model to dynamically balance attention across timesteps and spatial locations. Grain boundaries are extracted from each input frame, producing ``boundary masks'' that are resized to match the latent spatial dimensions. Prior to softmax normalization, attention scores at grain positions are suppressed by adding large negative values, putting all attention weight on the boundaries of the grains where evolution occurs. 
\begin{algorithm}[htbp]
\caption{Boundary mask extraction}
\label{alg:boundary_extraction}
\begin{algorithmic}[1]
\Statex \textbf{Input:}
\Statex \hspace{\algorithmicindent} $\bullet$ Microstructure sequence $\mathbf{X} = \{I_1, I_2, \ldots, I_T\}$, each $I_t \in \mathbb{R}^{H \times W}$ (grayscale image)
\Statex \hspace{\algorithmicindent} $\bullet$ Latent spatial dimensions $H' \times W'$
\Statex \textbf{Output:}
\Statex \hspace{\algorithmicindent} $\bullet$ Boundary mask sequence $\mathbf{M} = \{M_1, M_2, \ldots, M_T\}$, each $M_t \in [0, 1]^{H' \times W'}$
\Statex \hspace{\algorithmicindent}\quad ($1$ = boundary, $0$ = grain)
\Statex
\For{each timestep $t = 1, \ldots, T$}
    \State $\tau \gets \text{OtsuThreshold}(I_t)$ \Comment{Compute binary threshold with Otsu's method}
    \State $\mathbf{M}_{\text{bin}} \gets \mathbf{1}[I_t > \tau]$ \Comment{Binarize: 1 = boundary, 0 = grain}
    \State $M_t \gets \text{Resize}(\mathbf{M}_{\text{bin}},\; H', W')$ \Comment{Resize the mask to latent spatial dimensions}
\EndFor
\State \Return $\mathbf{M}$
\end{algorithmic}
\end{algorithm}

\begin{algorithm}[htbp]
\caption{Boundary-masked attention}
\label{alg:boundary_masked_attention}
\begin{algorithmic}[1]
\Statex \textbf{Input:}
\Statex \hspace{\algorithmicindent} $\bullet$ Encoder hidden states $\mathbf{S} \in \mathbb{R}^{T \times C \times H' \times W'}$
\Statex \hspace{\algorithmicindent} $\bullet$ Decoder hidden state $\mathbf{h}_{\text{dec}} \in \mathbb{R}^{C \times H' \times W'}$
\Statex \hspace{\algorithmicindent} $\bullet$ Boundary mask $\mathbf{M} \in [0, 1]^{T \times H' \times W'}$ (from Algorithm~\ref{alg:boundary_extraction})
\Statex \hspace{\algorithmicindent} $\bullet$ Learned projections $\mathbf{W}_{\text{enc}}, \mathbf{W}_{\text{dec}}, \mathbf{V}$; learned temperature $\tau_{\text{temp}}$; mask strength $\lambda \in [0, 1]$
\Statex \textbf{Output:}
\Statex \hspace{\algorithmicindent} $\bullet$ Context vector $\mathbf{c} \in \mathbb{R}^{C \times H' \times W'}$
\Statex \hspace{\algorithmicindent} $\bullet$ Attention weights $\boldsymbol{\alpha} \in \mathbb{R}^{T \times H' \times W'}$
\Statex
\Statex \textit{// Compute raw attention scores}
\State $\mathbf{K} \gets \mathbf{W}_{\text{enc}}(\mathbf{S})$ \Comment{Key: project encoder states, $T \times d_a \times H' \times W'$}
\State $\mathbf{Q} \gets \mathbf{W}_{\text{dec}}(\mathbf{h}_{\text{dec}})$ \Comment{Query: project decoder state, $1 \times d_a \times H' \times W'$}
\State $\mathbf{e} \gets \mathbf{V}\!\left(\tanh(\mathbf{K} + \mathbf{Q})\right)$ \Comment{Additive attention scores: $T \times H' \times W'$}
\State $\mathbf{e} \gets \mathbf{e} \;/\; \tau_{\text{temp}}$ \Comment{Temperature scaling}
\Statex \textit{// Apply boundary mask before softmax}
\State $\mathbf{e} \gets \mathbf{e} \odot \mathbf{M} + (-10^{4} \cdot \lambda) \cdot (1 - \mathbf{M})$ \Comment{Suppress grain scores}
\Statex \textit{// Normalize over the spatiotemporal space}
\State $\boldsymbol{\alpha} \gets \text{softmax}\!\left(\text{flatten}(\mathbf{e})\right)$ \Comment{Flatten to $(T \!\cdot\! H' \!\cdot\! W')$, then softmax}
\State $\boldsymbol{\alpha} \gets \text{reshape}(\boldsymbol{\alpha},\; T \times H' \times W')$
\Statex \textit{// Compute context vector as weighted sum over values}
\State $\mathbf{c} \gets \sum_{t,h,w} \boldsymbol{\alpha}_{t,h,w} \cdot \mathbf{S}_{t,:,h,w}$ \Comment{Value: encoder states $\mathbf{S}$; output: $C$}
\State $\mathbf{c} \gets \text{expand}(\mathbf{c},\; C \times H' \times W')$ \Comment{Broadcast to spatial dimensions}
\State \Return $\mathbf{c},\; \boldsymbol{\alpha}$
\end{algorithmic}
\end{algorithm}

As detailed in Algorithms~\ref{alg:boundary_extraction} and~\ref{alg:boundary_masked_attention}, and illustrated in Figure~\ref{fig:architecture}(b), the boundary-masked attention extends the baseline architecture by retaining all encoder ConvLSTM hidden states as spatiotemporal memory and incorporating the boundary-masked attention mechanism between the encoder and decoder ConvLSTM stages. At each autoregressive decoding step, the resulting context vector is concatenated with the decoder ConvLSTM input, allowing the model to access boundary information from any input timestep rather than relying solely on the compressed final encoder state.

In addition to its role in localizing relevant features, the proposed mechanism introduces an inductive bias aligned with the physics of grain boundary migration in grain growth. By restricting attention to the boundaries of the grains, the model is guided to learn relationships between boundary configurations at different timesteps, which is the information most relevant for predicting subsequent grain topology. This physics-informed constraint reduces the effective search space for the attention mechanism and potentially improves training efficiency, while ensuring that the learned attention patterns remain interpretable in terms of the underlying mechanism of grain growth. The attention weights for each test case in this study are visualized as heatmaps overlaid on the microstructure images, providing insight into the regions where the mechanism learns to focus given the boundary mask constraint.

\begin{table}[htbp]
\centering
\footnotesize
\renewcommand{\arraystretch}{1.2}
\setlength{\extrarowheight}{2pt}
\begin{tabularx}{\textwidth}{Xc}
\toprule
\textbf{Parameter} & \textbf{Value} \\
\midrule
Learning rate & $1 \times 10^{-4}$ \\ \cline{1-2}
Optimizer & AdamW \\ \cline{1-2}
Temporal window size (\si{\minute}) (Input--Predict) & $10$--$10$\\ \cline{1-2}
Dataset split ratio (\si{\percent}) (Train/Val/Test) & $70/20/10$ \\ \cline{1-2}
Batch size & $1$ \\ \cline{1-2}
Epochs & $60$ \\ \cline{1-2}
Gradient clipping threshold & $1.0$ \\
\bottomrule
\end{tabularx}
\caption{Hyperparameters used for model training.}
\label{tab:training_hyperparams}
\end{table}

In terms of model complexity, the integration of the attention mechanism resulted in a modest increase compared to the baseline. The boundary-masked attention model contains \num{1.182}M parameters compared to \num{1.174}M parameters in the baseline model, representing only a \SI{0.70}{\percent} increase. This increase arises from the learned projection weights for query, key, and value computations. Since architectural modifications were required, the boundary-masked attention model was trained from scratch on the same in-distribution synthetic dataset used for the baseline model, namely the \num{647} TRM-generated sequences of pure grain growth with physical parameters described in Section~\ref{sec:baseline_model}. Training was conducted on an NVIDIA A100 \SI{40}{\giga\byte} GPU using identical hyperparameters, as listed in Table~\ref{tab:training_hyperparams}. As with the baseline model, the weights of the boundary-masked attention model were kept fixed throughout all OOD evaluations.

\subsection{Experimental Setup}
\label{sec:experimental_setup}

To systematically assess the generalization capability of both the baseline and the boundary-masked attention models, a comprehensive experimental framework was designed consisting of three distinct OOD test cases, each representing a different type of distribution shift relevant to practical industrial applications, as detailed in Sections~\ref{sec:test_case_1}--\ref{sec:test_case_3}. By evaluating both models under identical conditions across all three test cases, their generalization performance could be directly compared and the conditions under which the attention mechanism provided benefits versus limited improvement could be identified.

\subsubsection{Comparison Protocol}

For all three test cases, both models were evaluated using an identical protocol to ensure fair comparison. Each test case used a validated TRM simulation tool to generate a complete evolution of microstructure from $t = \SI{0}{\minute}$ to $t = \SI{59}{\minute}$ under isothermal annealing at \SI{1325.15}{\kelvin}, producing \num{60} frames per sequence at $1024 \times 1024$ pixel resolution. The first \SI{10}{\minute} of evolution ($t = \SI{0}{\minute}$ to $t = \SI{9}{\minute}$) served as model input, consistent with the S-10-10 configuration, while the remaining frames ($t = \SI{10}{\minute}$ to $t = \SI{59}{\minute}$) served as ground truth for evaluation. Predictions were generated autoregressively in \SI{10}{\minute} increments, with each predicted window used as input to the subsequent one without access to ground truth.

\subsubsection{Test Case 1: Experimental Microstructures}
\label{sec:test_case_1}

\begin{figure}[h!]
    \centering
    \begin{subfigure}[t]{0.48\linewidth}
    \centering
    \includegraphics[width=\linewidth]{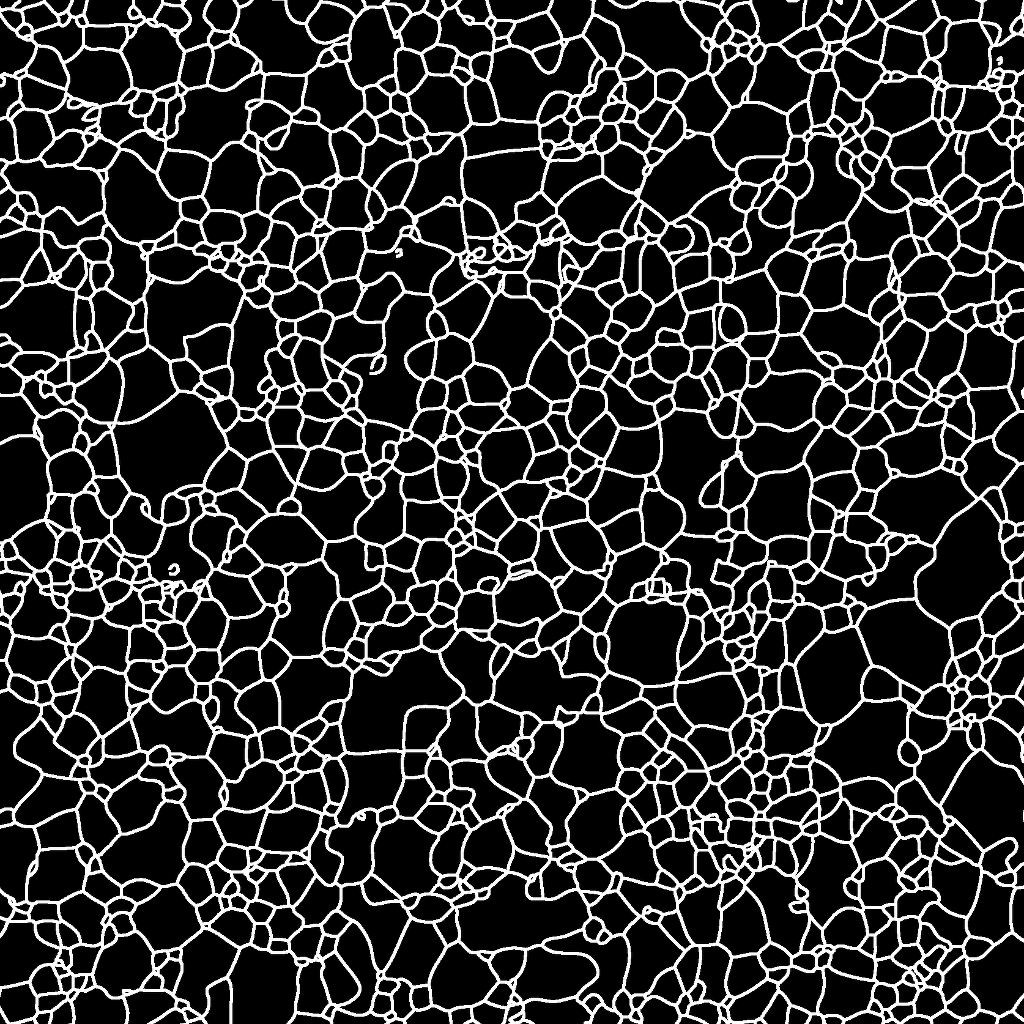}
    \caption{}
    \label{fig:test_case_1_init_0}
    \end{subfigure}
    \hfill
    \begin{subfigure}[t]{0.48\linewidth}
    \centering
    \includegraphics[width=\linewidth]{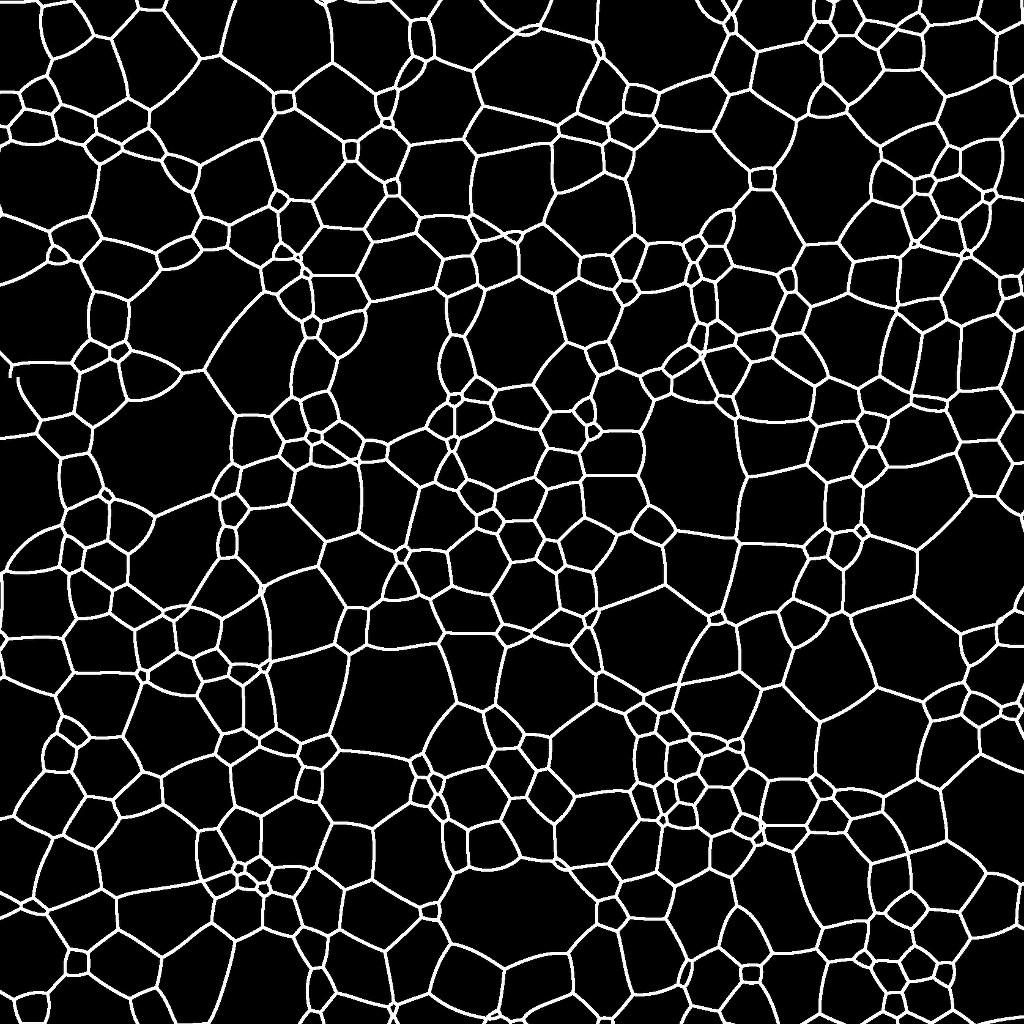}
    \caption{}
    \label{fig:test_case_1_init_59}
    \end{subfigure}
    \caption{Microstructure images for Test Case 1 (experimental microstructures): (a) initial state ($t = \SI{0}{\minute}$), containing approximately \num{1800} grains with rough and irregular grain boundaries characteristic of experimental imaging; (b) final state ($t = \SI{59}{\minute}$) after isothermal annealing at \SI{1325.15}{\kelvin}, with a reduced mobility representative of 304L stainless steel ($\mu_\gamma \approx \SI{e-6}{\milli\meter\squared\per\second}$).}
    \label{fig:test_case_1}
\end{figure}

The first test case evaluated generalization of both models to experimental microstructures, illustrated in Figure~\ref{fig:test_case_1}(a,b). While the overall grain morphology resembles that of the synthetic training data, with both exhibiting polycrystalline structures with comparable grain sizes and topological arrangements, the grain boundaries of this experimental microstructure differ substantially from those in the synthetic training data. Synthetic microstructures in the training dataset typically exhibit smooth grain boundaries, whereas the boundaries of the experimental microstructure exhibit roughness and irregularities inherent to real materials characterization.

In terms of statistical characteristics, the initial microstructure contains approximately \num{1800} grains, with a mean value of surface-weighted Equivalent Circle Radius (ECR) distribution ($\overline{R}$) of \SI{38}{\micro\meter}. After one hour of thermal treatment, the microstructure contains approximately \num{700} grains with an $\overline{R}$ of \SI{48}{\micro\meter} at the final state. This evolution is consistent with pure grain growth coarsening behavior, with the $\overline{R}$ increasing by approximately \SI{26}{\percent} and the total number of grains reducing by approximately \SI{61}{\percent}. The ECR and grain neighbor count distributions at the initial state ($t = \SI{0}{\minute}$) are shown in Figure~\ref{fig:initial_dist_overview}(a,b).

Morphologically, the roughness of grain boundaries in this test case presented a fundamental challenge for the models trained exclusively on synthetic data generated using LavoGen, due to the fact that the models had to predict the migration of irregular boundaries despite having learned exclusively from smooth, idealized boundaries during training. In grain growth, boundary roughness evolves toward smoother morphologies as growth proceeds. However, the initial roughness can still impact early predictions and cause error accumulation over time. Furthermore, no preprocessing methods such as boundary smoothing were applied to the input data. The models processed the raw images with no preprocessing applied, ensuring a realistic assessment of generalization capability. The ability of the models to generalize to this type of real-world data is critical for practical applications, as it reflects conditions encountered in industrial settings where microstructure characterization often involves inherent noise and imperfections.

\subsubsection{Test Case 2: Synthetic Bimodal Microstructures}
\label{sec:test_case_2}

\begin{figure}[h!]
    \centering
    \begin{subfigure}[t]{0.48\linewidth}
    \centering
    \includegraphics[width=\linewidth]{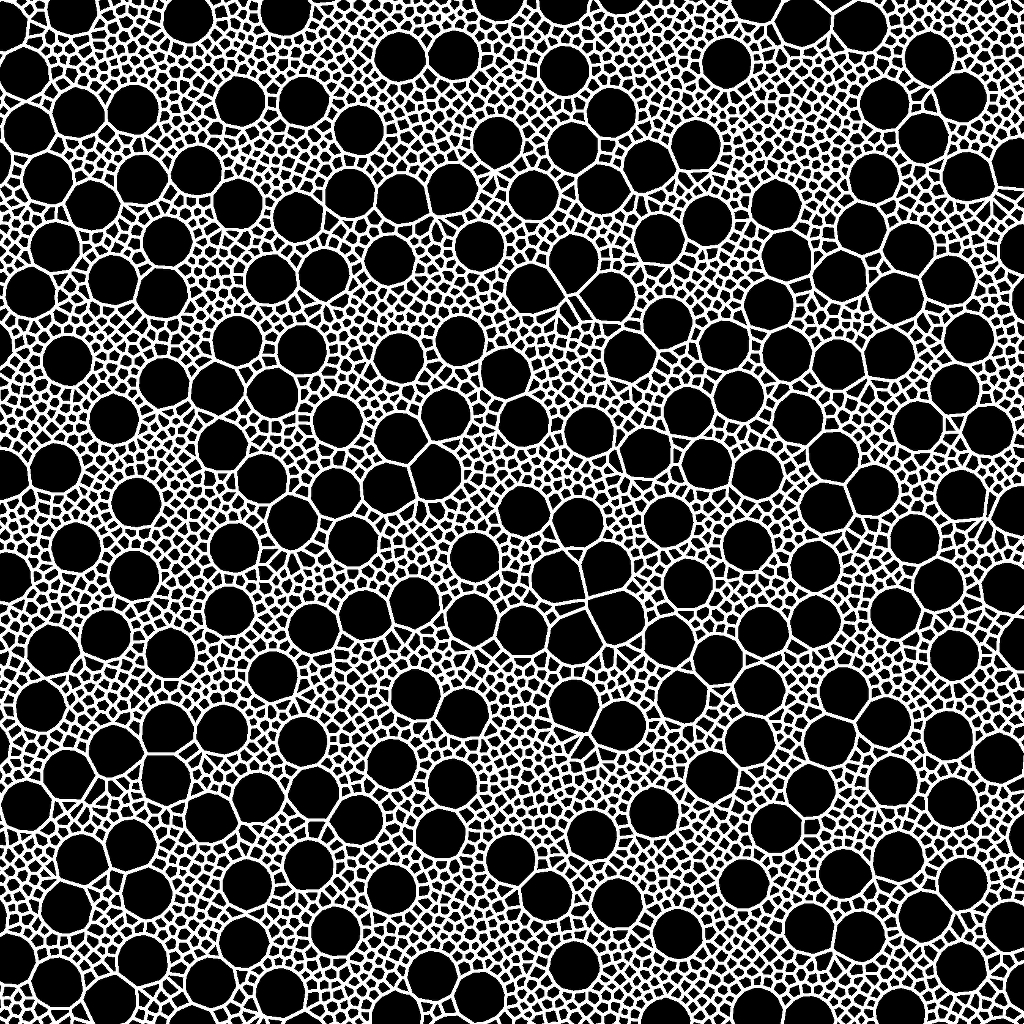}
    \caption{}
    \label{fig:test_case_2_init_0}
    \end{subfigure}
    \hfill
    \begin{subfigure}[t]{0.48\linewidth}
    \centering
    \includegraphics[width=\linewidth]{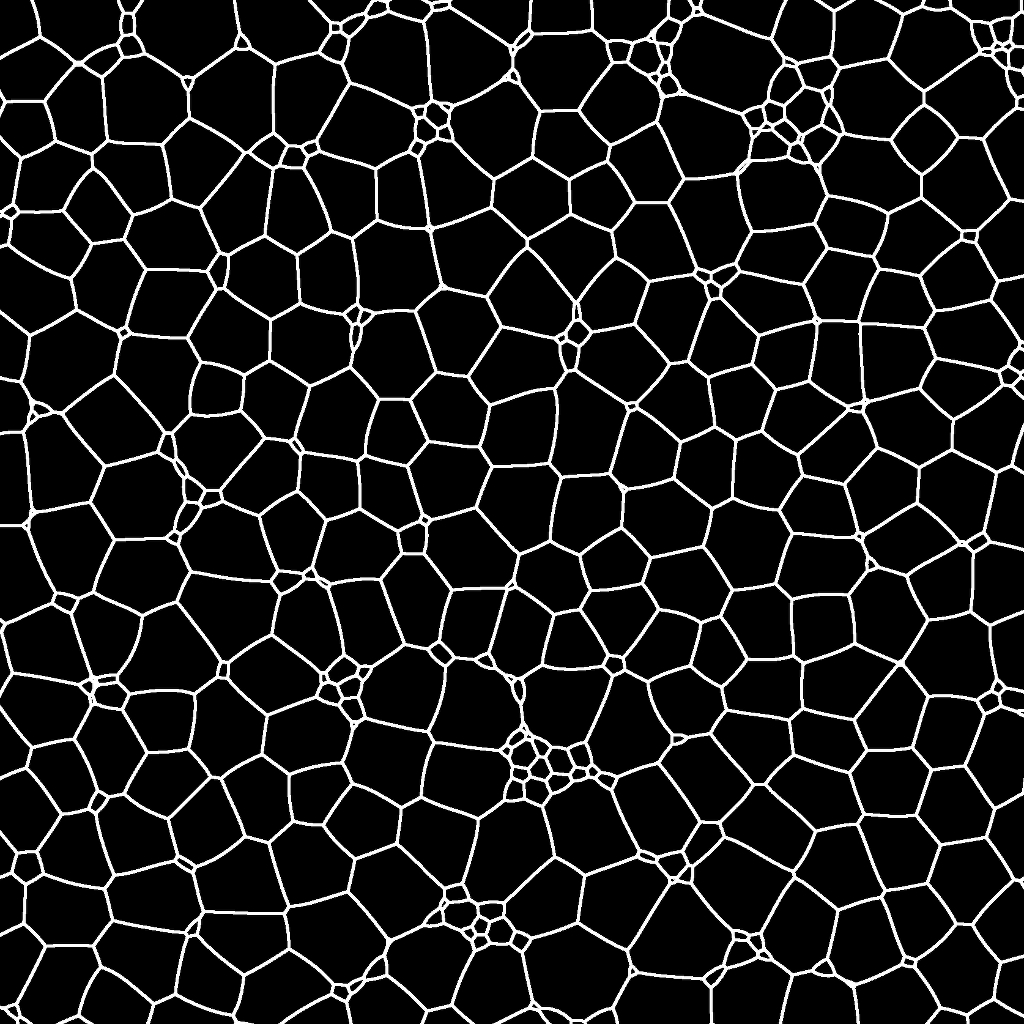}
    \caption{}
    \label{fig:test_case_2_init_59}
    \end{subfigure}
    \caption{Microstructure images for Test Case 2 (synthetic bimodal microstructures): (a) initial state ($t = \SI{0}{\minute}$), containing approximately \num{3500} grains with a bimodal ECR distribution; (b) final state ($t = \SI{59}{\minute}$) after isothermal annealing at \SI{1325.15}{\kelvin}.}
    \label{fig:test_case_2}
\end{figure}

The second test case employed synthetic data generated using the TRM simulation approach as the training data, yet with the grain size distribution shifted from the Gaussian form present in the training data to a bimodal form. The initial and final states of the microstructure are visualized in Figure~\ref{fig:test_case_2}(a,b).

Quantitatively, the bimodal microstructure was generated using the TRM simulations with initial conditions designed to produce two distinct grain size populations. At the initial state, the microstructure consisted of approximately \num{3500} grains with an overall $\overline{R}$ of \SI{20}{\micro\meter}. After one hour of thermal treatment, the final state contained approximately \num{500} grains with an $\overline{R}$ of \SI{50}{\micro\meter}. This resulted in a $\overline{R}$ increase of \SI{150}{\percent} while the total number of grains reduced by approximately \SI{86}{\percent}, reflecting aggressive coarsening of the fine-grained population. The ECR and grain neighbor count distributions at the initial state ($t = \SI{0}{\minute}$) are shown in Figure~\ref{fig:initial_dist_overview}(c,d).

Fundamentally, the key challenge in this test case lies in the distribution shape. The initial grain size distribution differs qualitatively from the training data, as the models were trained only on unimodal Gaussian grain size distributions while this test case features two distinct peaks, as shown in Figure~\ref{fig:initial_dist_overview}(c). The trained models had to extrapolate beyond the learned parameter space to predict the evolution of this bimodal distribution. As evolution proceeded, smaller grains were preferentially eliminated through boundary migration while larger grains continued to grow, producing a gradual transition from a bimodal toward a normal distribution, as illustrated in Figure~\ref{fig:test_case_2}(a,b). This test case therefore assessed whether the models captured the underlying physics of curvature-driven boundary migration or only memorized patterns specific to unimodal Gaussian grain size statistics.

\subsubsection{Test Case 3: Abnormal Grain Growth}
\label{sec:test_case_3}

\begin{figure}[h!]
    \centering
    \begin{subfigure}[t]{0.48\linewidth}
    \centering
    \includegraphics[width=\linewidth]{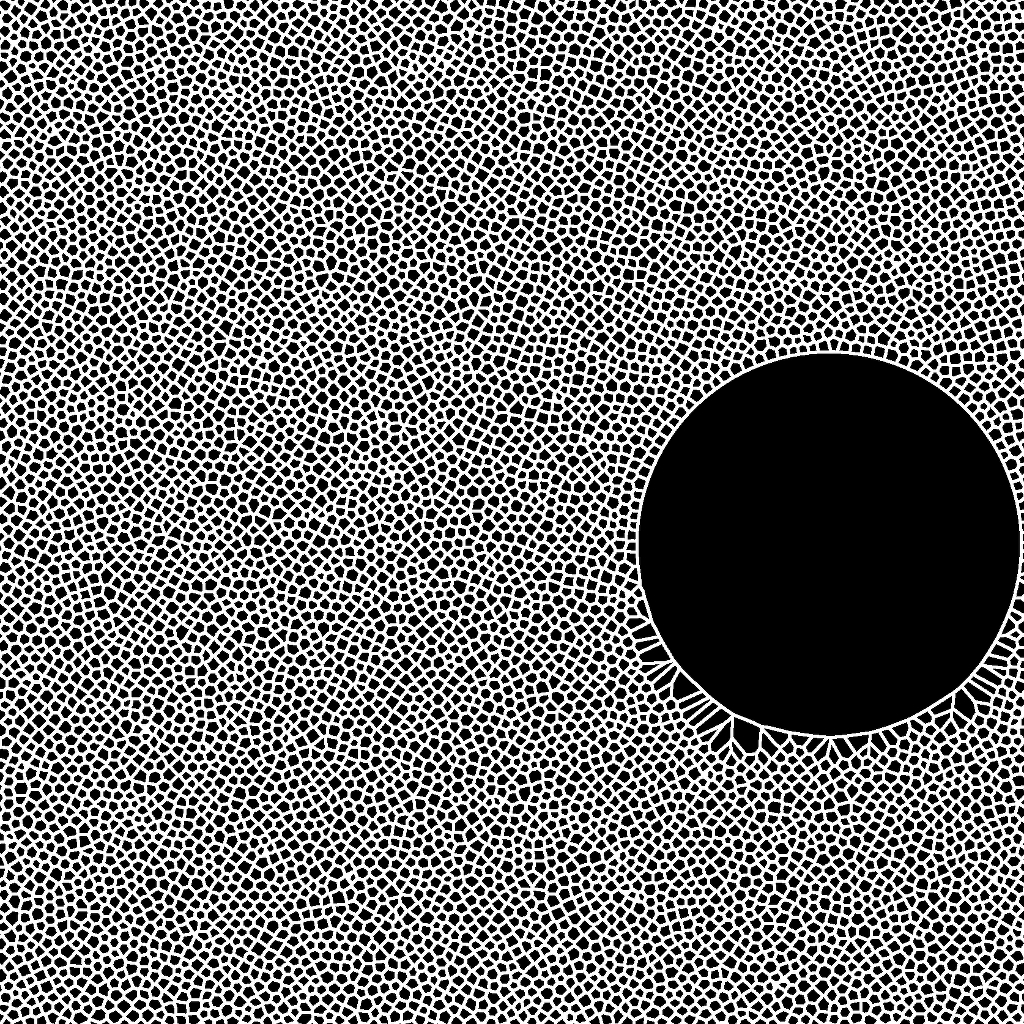}
    \caption{}
    \label{fig:test_case_3_init_0}
    \end{subfigure}
    \hfill
    \begin{subfigure}[t]{0.48\linewidth}
    \centering
    \includegraphics[width=\linewidth]{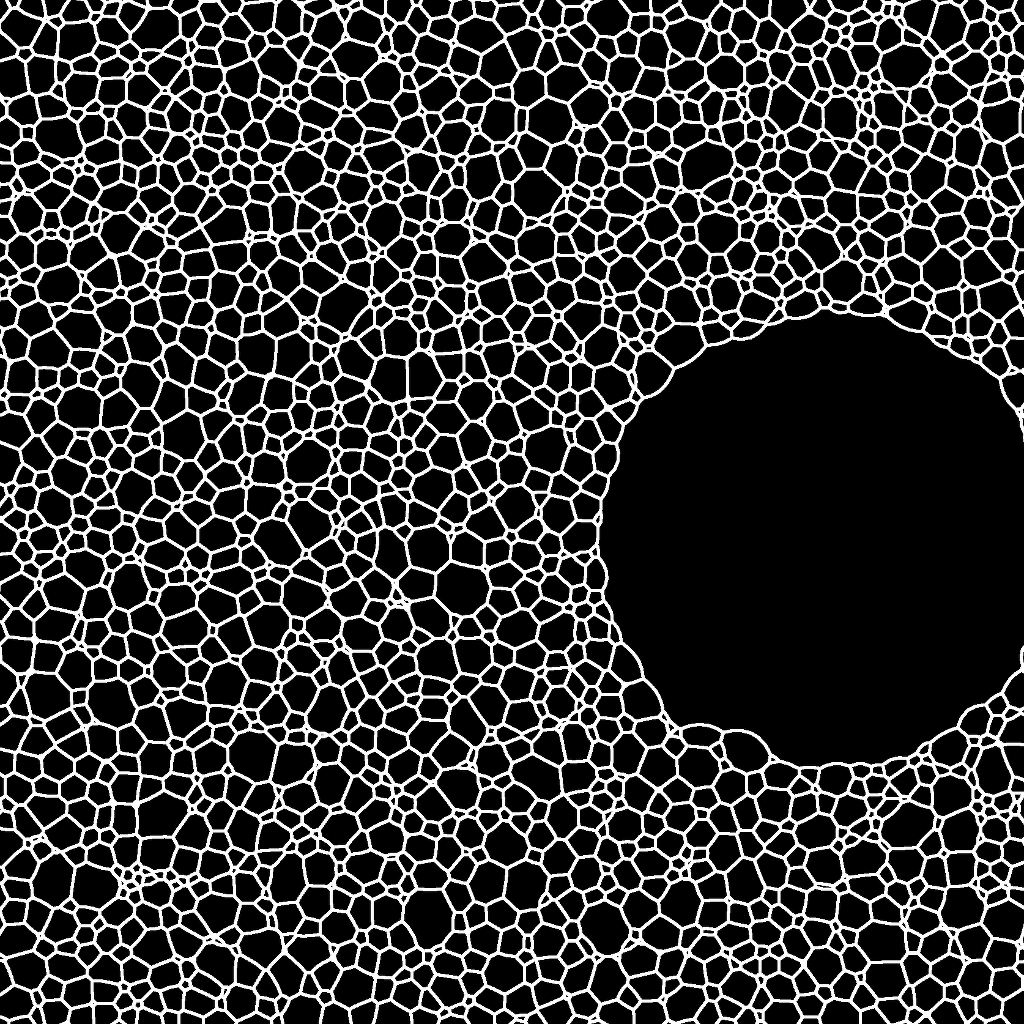}
    \caption{}
    \label{fig:test_case_3_init_59}
    \end{subfigure}
    \caption{Microstructure images for Test Case 3 (abnormal grain growth): (a) initial state ($t = \SI{0}{\minute}$), containing approximately \num{6000} grains with a single abnormal grain; (b) final state ($t = \SI{59}{\minute}$), showing the abnormal grain having consumed approximately \SI{20}{\percent} of the total domain area.}
    \label{fig:test_case_3}
\end{figure}

The final generalization test case investigated the ability of the models to predict abnormal grain evolution. This test case represented a fundamental difference in physical mechanism compared to the regular grain growth used during training. In the first two test cases, all grains evolve with comparable boundary migration kinetics determined primarily by local characteristics, resulting in relatively uniform coarsening kinetics across the microstructure. In contrast, abnormal grain growth involves one or more large ``abnormal'' grains that progressively consume the surrounding smaller ``normal'' grains~\cite{krill2023extreme, rollett2017recrystallization}. In this specific simulation, the initial advantage of the abnormal grain arises solely from its size, while its boundary mobility is identical to that of all other grain boundaries in the microstructure.

\begin{figure}[h!]
    \centering
    \begin{subfigure}[b]{0.48\linewidth}
        \centering
        \includegraphics[width=\linewidth]{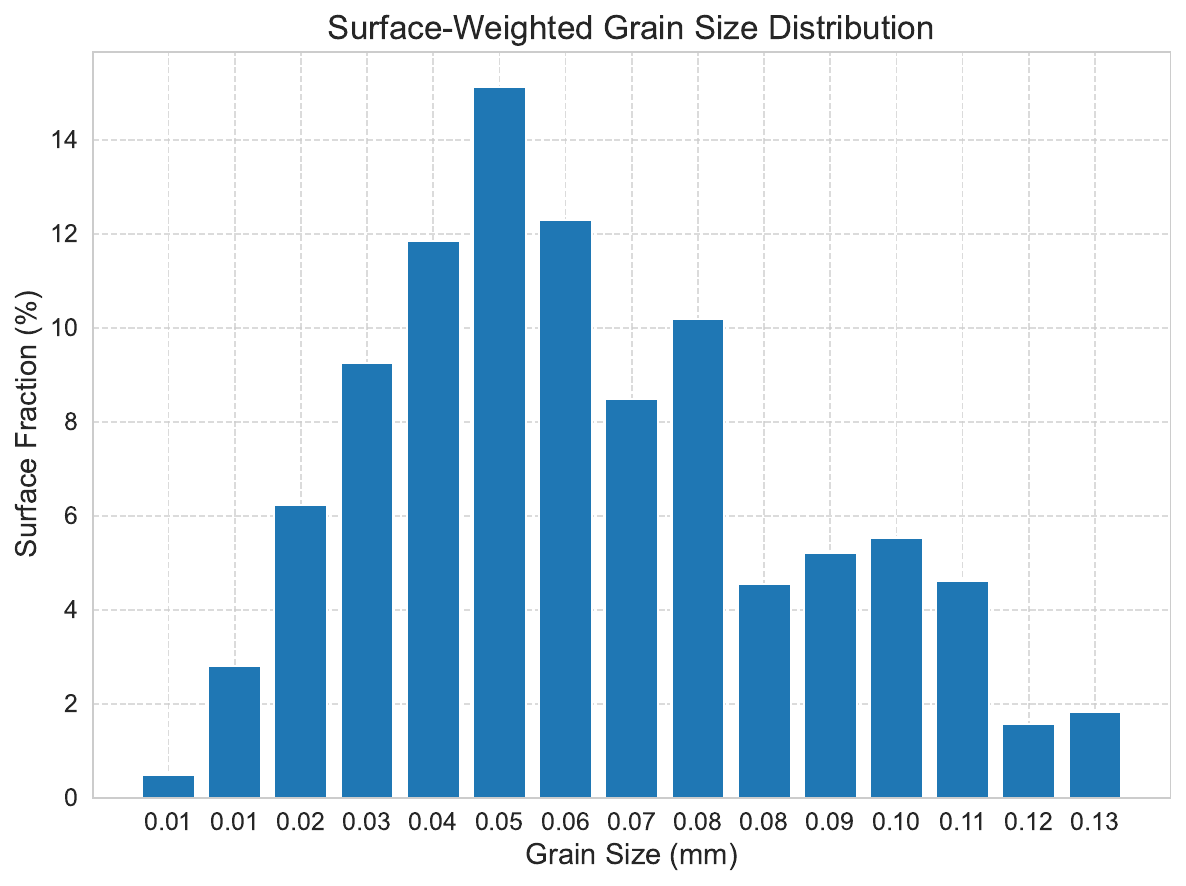}
        \caption{}
        \label{fig:initial_dist_tc1_ecr}
    \end{subfigure}
    \hfill
    \begin{subfigure}[b]{0.48\linewidth}
        \centering
        \includegraphics[width=\linewidth]{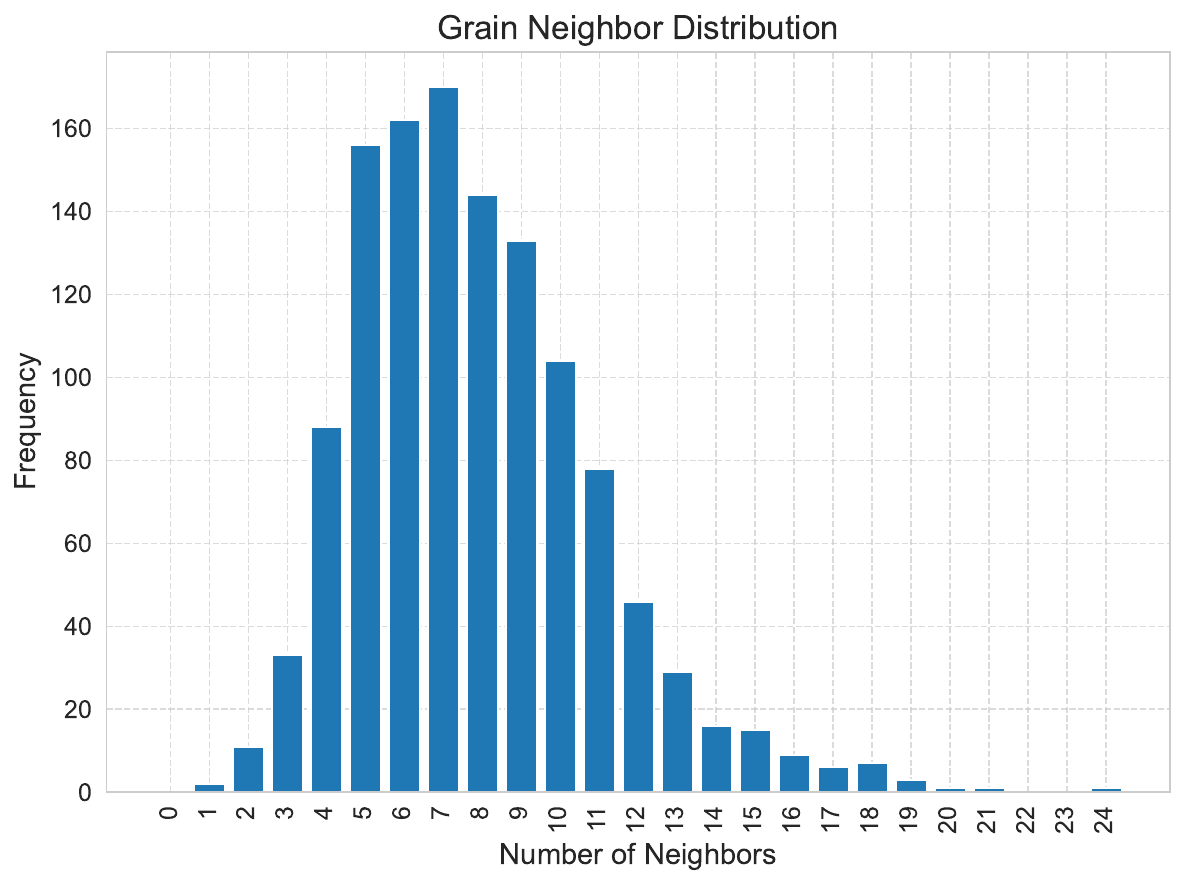}
        \caption{}
        \label{fig:initial_dist_tc1_neighbour}
    \end{subfigure}

    \vspace{1em}

    \begin{subfigure}[b]{0.48\linewidth}
        \centering
        \includegraphics[width=\linewidth]{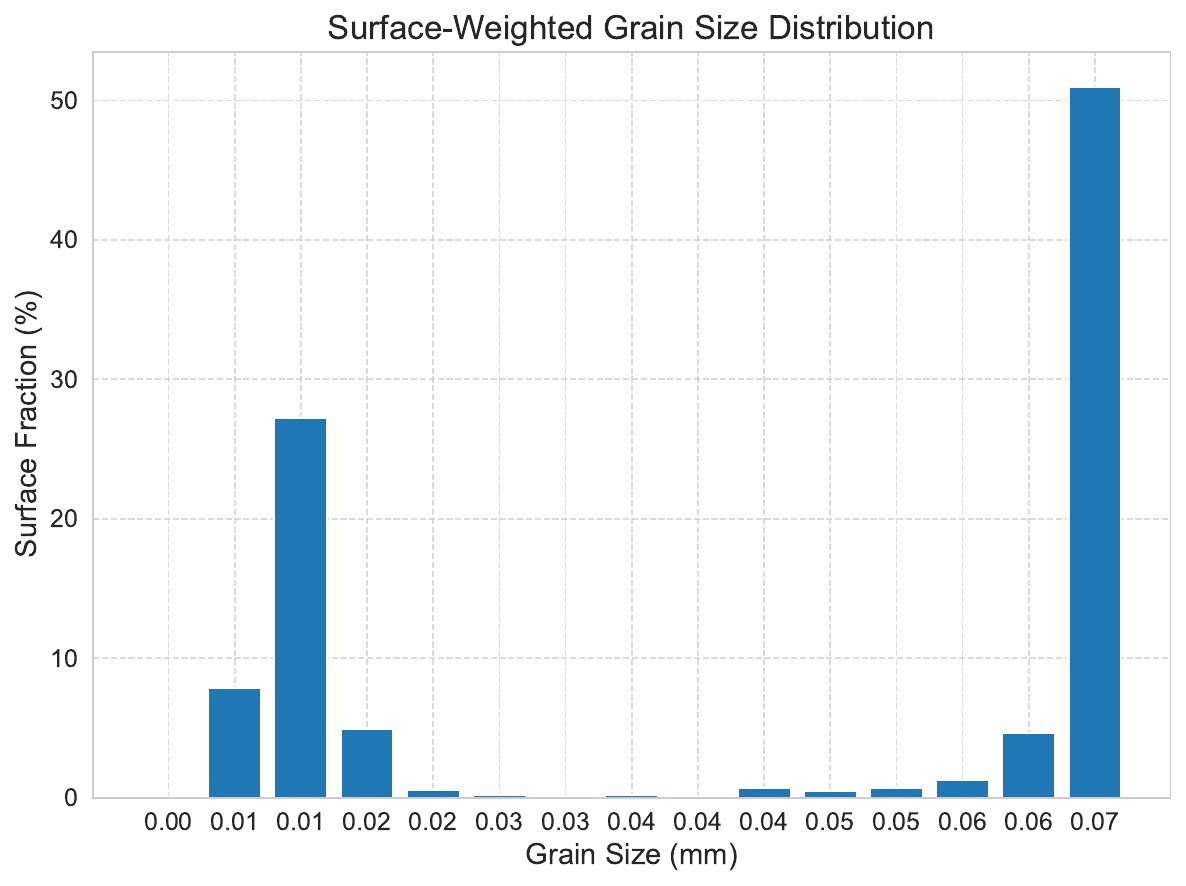}
        \caption{}
        \label{fig:initial_dist_tc2_ecr}
    \end{subfigure}
    \hfill
    \begin{subfigure}[b]{0.48\linewidth}
        \centering
        \includegraphics[width=\linewidth]{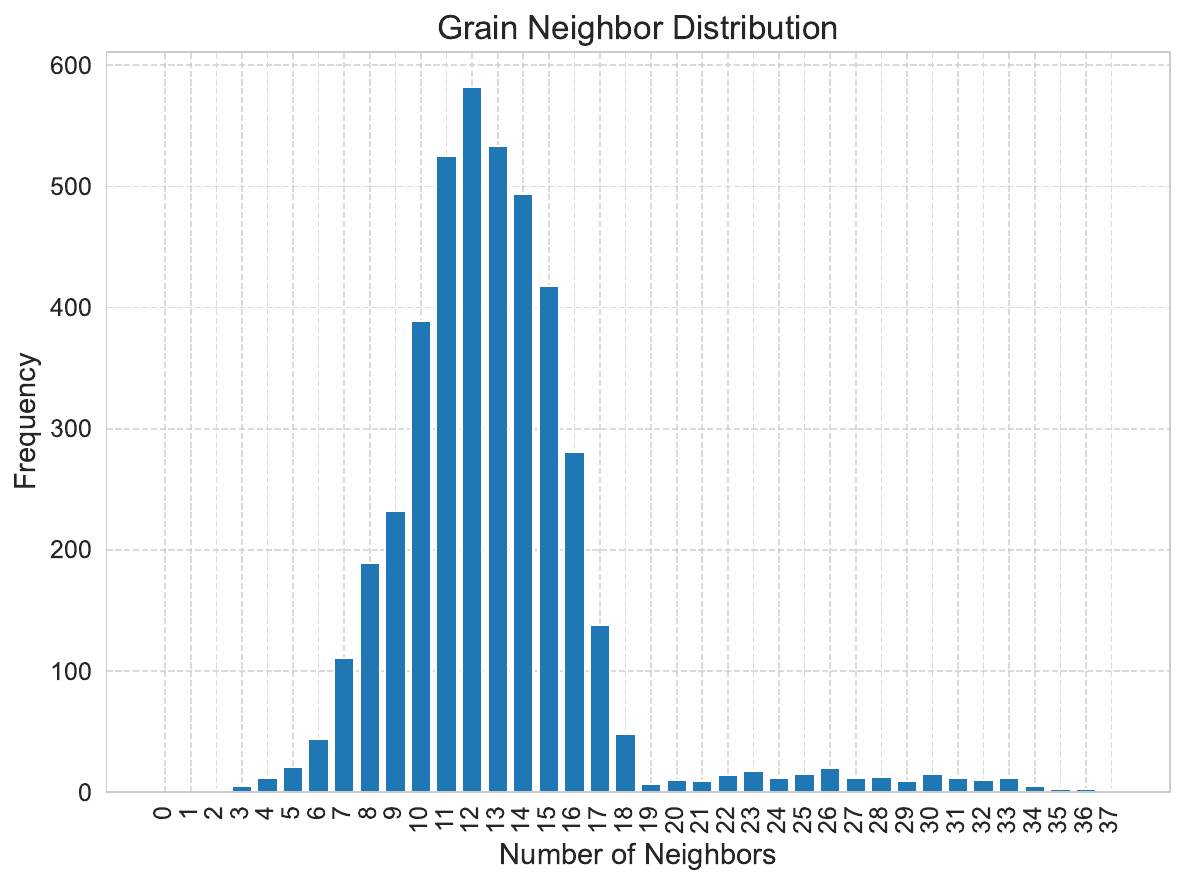}
        \caption{}
        \label{fig:initial_dist_tc2_neighbour}
    \end{subfigure}

    \vspace{1em}

    \begin{subfigure}[b]{0.48\linewidth}
        \centering
        \includegraphics[width=\linewidth]{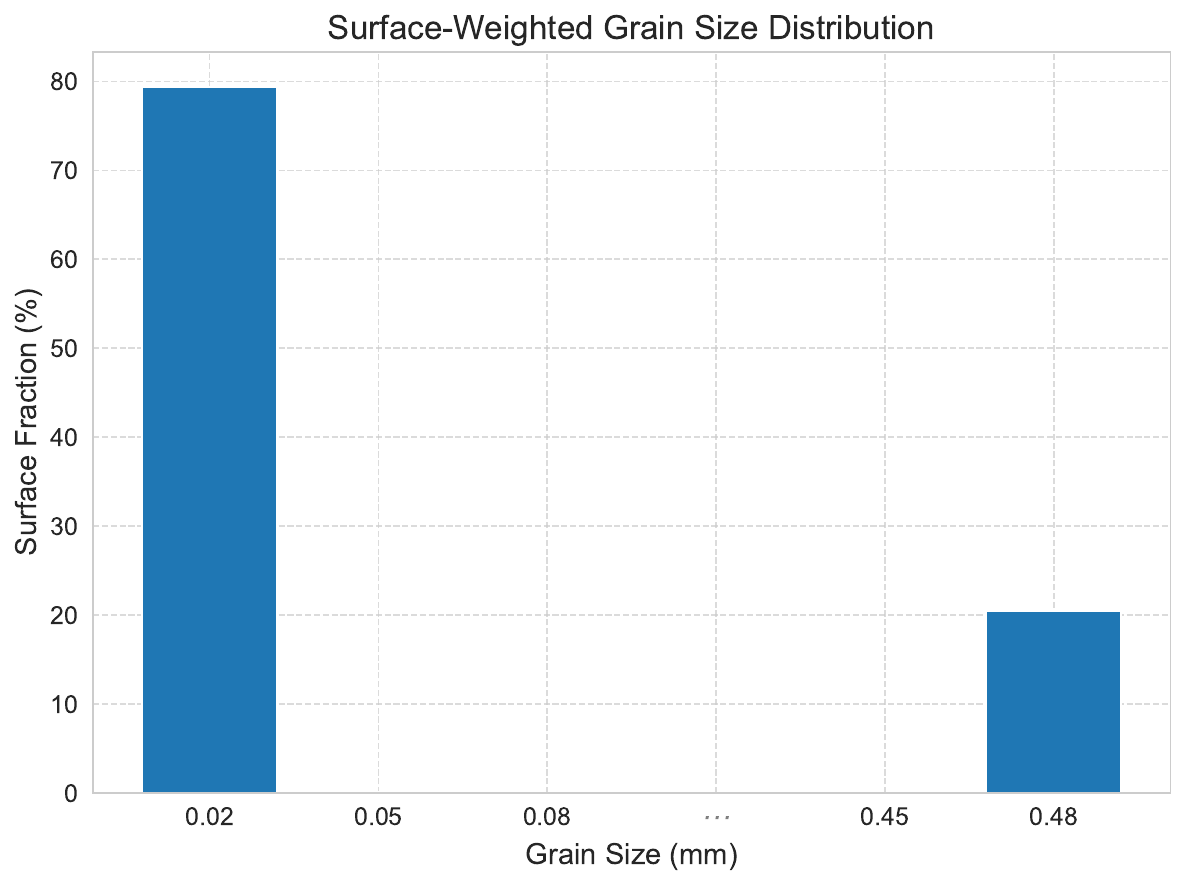}
        \caption{}
        \label{fig:initial_dist_tc3_ecr}
    \end{subfigure}
    \hfill
    \begin{subfigure}[b]{0.48\linewidth}
        \centering
        \includegraphics[width=\linewidth]{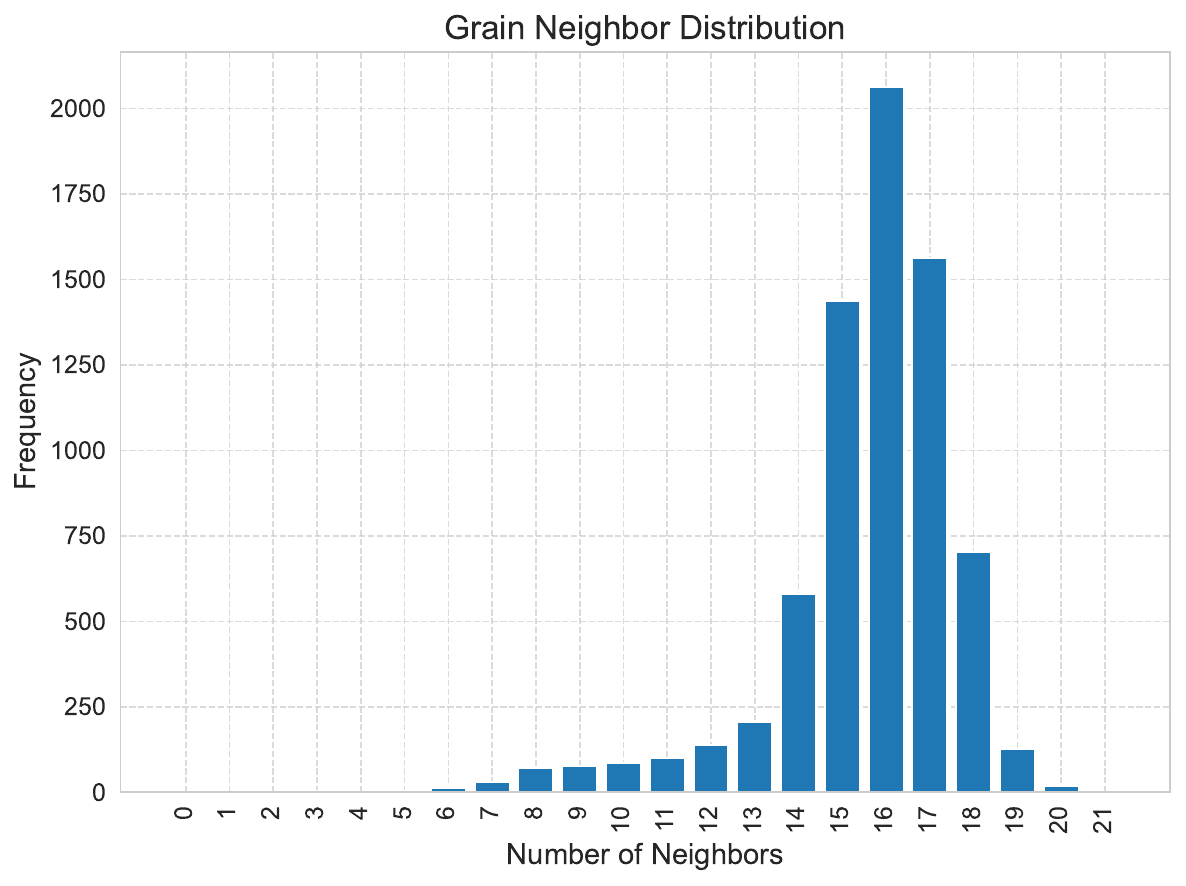}
        \caption{}
        \label{fig:initial_dist_tc3_neighbour}
    \end{subfigure}

    \caption{Initial state distributions ($t = \SI{0}{\minute}$) for all three OOD test cases, where left panels show surface-weighted ECR distributions and right panels show grain neighbor count distributions: (a, b) Test Case 1 (experimental microstructures); (c, d) Test Case 2 (synthetic bimodal microstructures); (e, f) Test Case 3 (abnormal grain growth).}
    \label{fig:initial_dist_overview}
\end{figure}

As illustrated in Figure~\ref{fig:test_case_3}(a,b), the test sequence featured a single abnormal grain surrounded by numerous smaller normal grains. At the initial state, the microstructure contained approximately \num{6000} grains with an $\overline{R}$ of \SI{12}{\micro\meter}. After one hour of evolution, the single abnormal grain reached an ECR of \SI{600}{\micro\meter} and consumed almost \SI{20}{\percent} of the total surface area of the microstructure. The total number of grains decreased to approximately \num{2000}, and the $\overline{R}$ increased to \SI{25}{\micro\meter} (a \SI{108}{\percent} increase), reflecting the topological changes as the abnormal grain progressively consumed its neighbors. The ECR and grain neighbor count distributions at the initial state ($t = \SI{0}{\minute}$) are shown in Figure~\ref{fig:initial_dist_overview}(e,f).

In this scenario, both models had to simultaneously predict two distinct types of boundary migration: moving boundaries between similar-sized normal grains and around the single abnormal grain. This multi-scale dynamic is absent from the training data, and this specific test case therefore assessed whether the models learned general principles of interface energy minimization, or whether they only memorized patterns specific to the uniform kinetics of pure grain growth.

\subsection{Evaluation Metrics}
\label{sec:evaluation_metrics}

The evaluation framework from the previous study~\cite{TEP2025121486} was adopted with one additional topological metric. Pixel-level and perceptual quality was assessed through Boundary-focused Mean Absolute Error ($\text{MAE}_{\text{b}}$)~\cite{TEP2025121486}, Boundary-focused Mean Squared Error ($\text{MSE}_{\text{b}}$)~\cite{TEP2025121486}, Peak Signal-to-Noise Ratio (PSNR)~\cite{SSIM2004}, and SSIM~\cite{SSIM2004}. Grain-level statistics were quantified from surface-weighted ECR distributions using Kullback-Leibler (KL) divergence~\cite{KLDivergence1951}, Wasserstein distance~\cite{Wasserstein2013}, and $\overline{R}$ error.

A further metric introduced in this study is the grain neighbor count distribution, which characterizes network connectivity by recording the number of boundary-sharing neighbors per grain. Two-dimensional grain structures under pure grain growth converge toward approximately six neighbors per grain on average at extended annealing times under isotropic grain growth conditions, following von Neumann--Mullins predictions~\cite{mullins1956two}, while bimodal distributions and abnormal grain growth can exhibit deviations. Distributional agreement is quantified through KL divergence and Wasserstein distance, enabling detection of topological prediction failures that may not be captured by the other metrics. Primary comparisons targeted the \SI{59}{\minute} final state, supplemented by temporal tracking at \SI{10}{\minute} intervals.

\section{Results}
\label{sec:results}

\subsection{Test Case 1: Experimental Microstructures}

\begin{figure}[h!]
    \centering
    \begin{subfigure}[t]{0.48\linewidth}
    \centering
    \includegraphics[width=\linewidth]{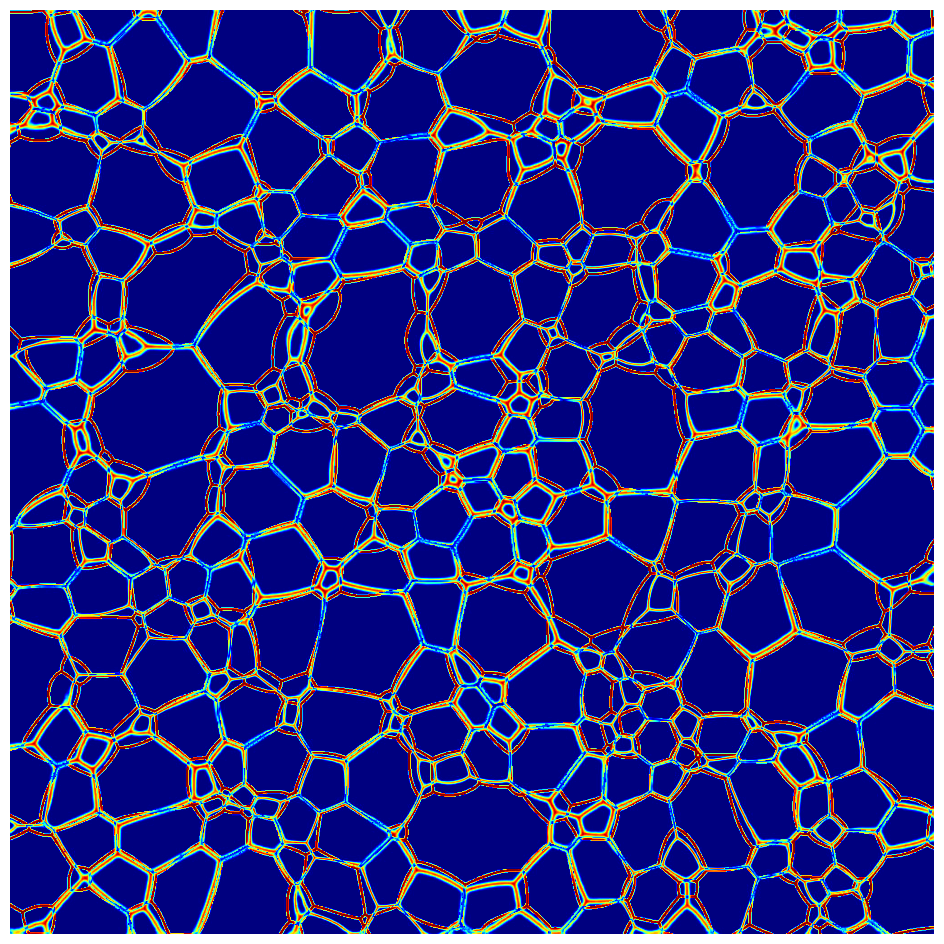}
    \caption{}
    \label{fig:result_case1_heatmap_a}
    \end{subfigure}
    \hfill
    \begin{subfigure}[t]{0.48\linewidth}
    \centering
    \includegraphics[width=\linewidth]{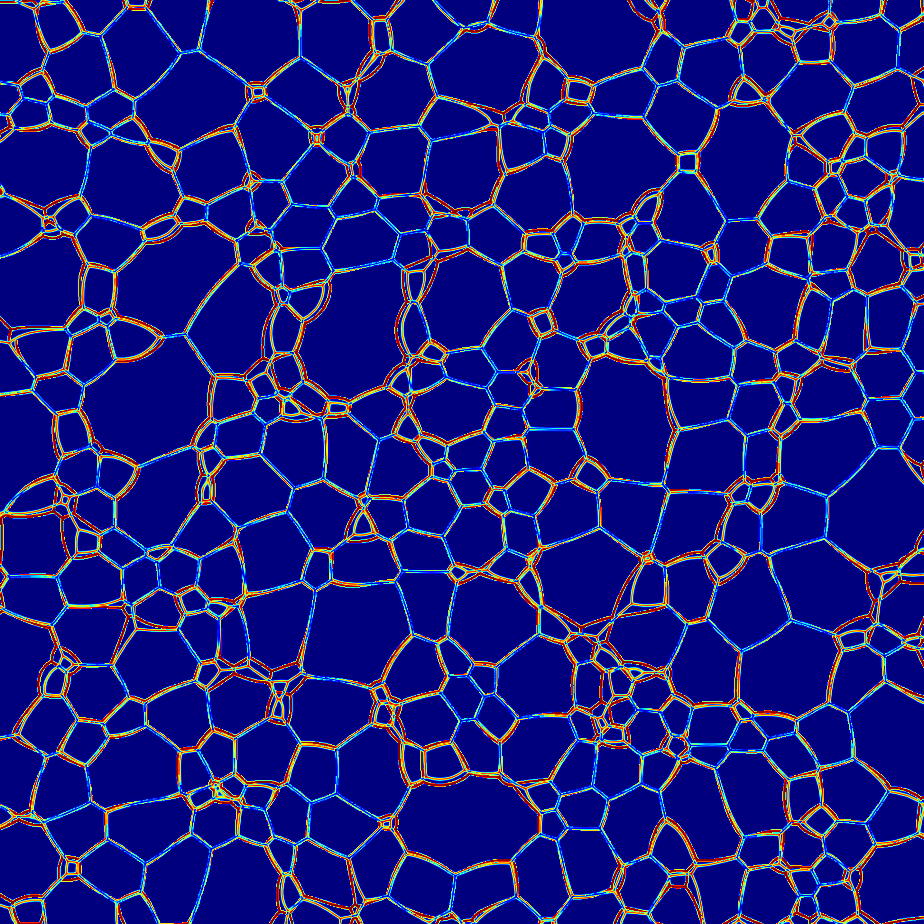}
    \caption{}
    \label{fig:result_case1_heatmap_b}
    \end{subfigure}
    \vspace{0.10cm}
    \centering
    \includegraphics[width=1\textwidth]{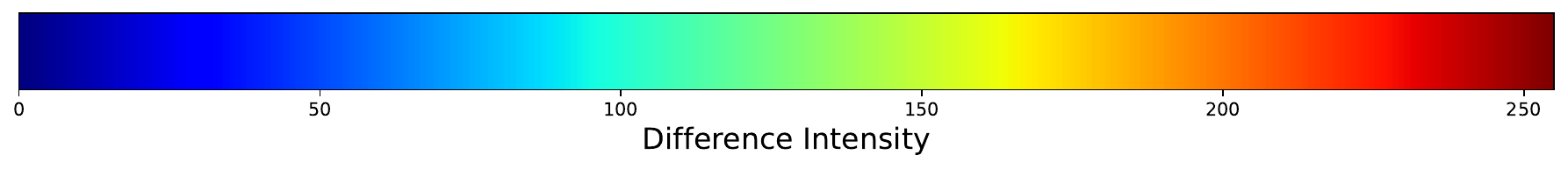}
    \caption{Error heatmap for Test Case 1 (experimental microstructures) at $t = \SI{59}{\minute}$: (a) baseline model; (b) attention model. (For interpretation of the references to color in this figure legend, the reader is referred to the web version of this article.)}
    \label{fig:result_case1_heatmap}
\end{figure}

\begin{table}[htbp]
    \centering
    \resizebox{\textwidth}{!}{%
    \begin{tabular}{l|cccc|ccc|cc}
    \toprule
    \multirow{2}{*}{\textbf{Model}} & \multicolumn{4}{c|}{\textbf{Pixel-wise and Perceptual}} & \multicolumn{3}{c|}{\textbf{ECR Distribution}} & \multicolumn{2}{c}{\textbf{Neighbor Count Distribution}} \\
    \cmidrule(lr){2-5} \cmidrule(lr){6-8} \cmidrule(lr){9-10}
    & $\text{MSE}_\text{b}$ $\downarrow$ & $\text{MAE}_\text{b}$ $\downarrow$ & PSNR (dB) $\uparrow$ & SSIM $\uparrow$ & $\overline{R}$ Err. (\%) $\downarrow$ & KL $\downarrow$ & W $\downarrow$ & KL $\downarrow$ & W $\downarrow$ \\
    \midrule
    Baseline & 0.3064 & 0.4729 & 9.50 & 0.5603 & 14.40 & 1.5049 & 0.0111 & 0.1679 & 0.5969 \\
    Boundary-Masked Attention & \textbf{0.2722} & \textbf{0.4339} & \textbf{10.701} & \textbf{0.6593} & \textbf{7.04} & \textbf{0.0604} & \textbf{0.0061} & \textbf{0.0313} & \textbf{0.1509} \\
    \bottomrule
    \end{tabular}%
    }

    \vspace{2pt}
    {\footnotesize KL = KL divergence (predicted $\rightarrow$ ground truth); W = Wasserstein distance. Best values are shown in \textbf{bold}.\par}
    \caption{Quantitative evaluation metrics for Test Case 1 (experimental microstructures) at $t = \SI{59}{\minute}$.}
    \label{tab:case1_metrics}
\end{table}

\begin{figure}[h!]
    \centering
    \begin{subfigure}[t]{0.48\linewidth}
    \centering
    \includegraphics[width=\linewidth]{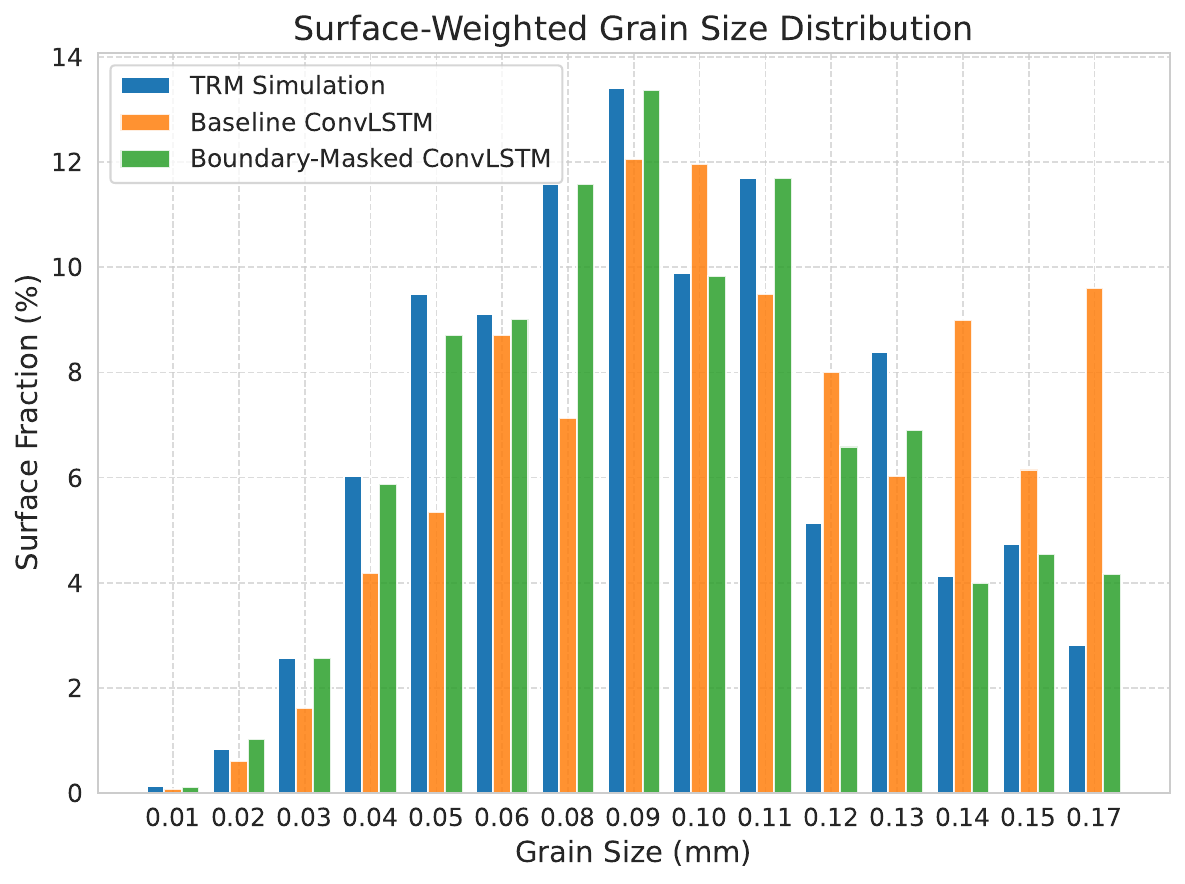}
    \caption{}
    \label{fig:result_case1_dist_a}
    \end{subfigure}
    \hfill
    \begin{subfigure}[t]{0.48\linewidth}
    \centering
    \includegraphics[width=\linewidth]{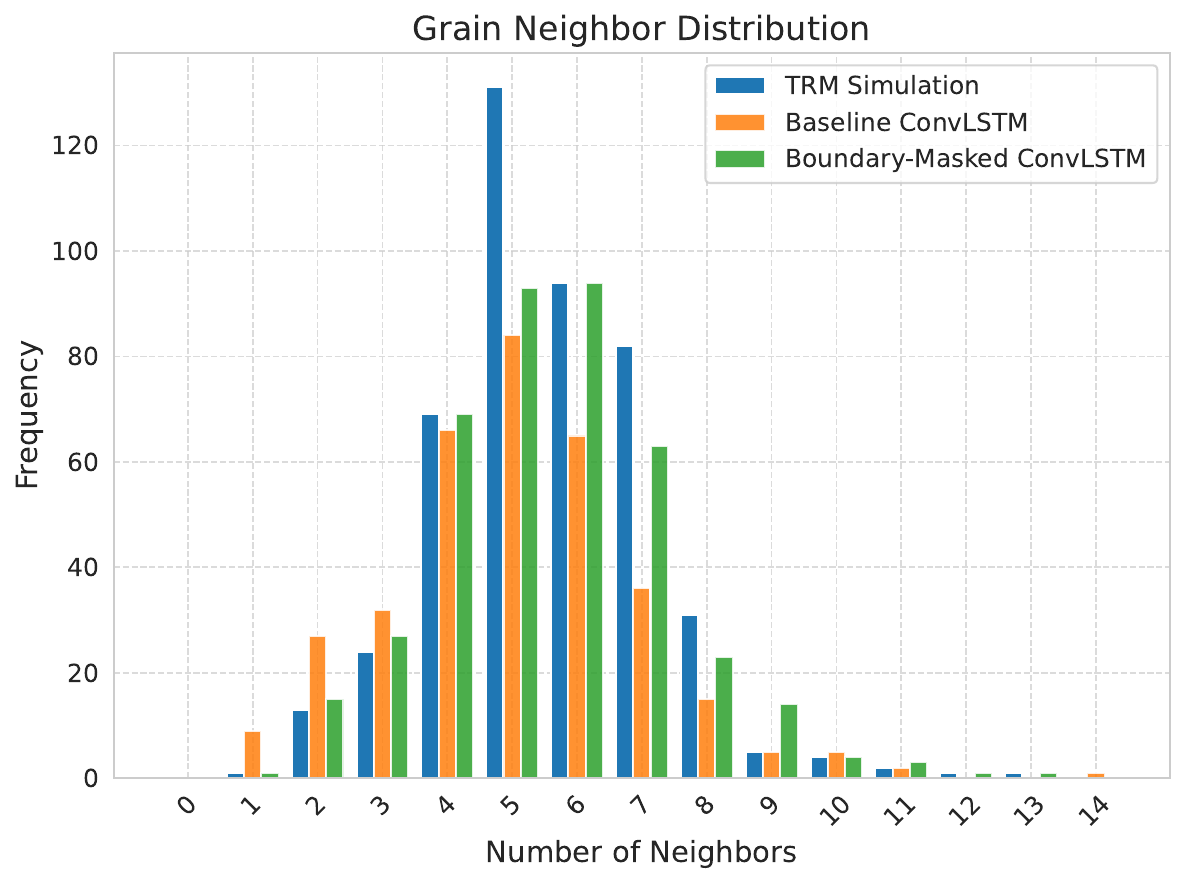}
    \caption{}
    \label{fig:result_case1_dist_b}
    \end{subfigure}
    \caption{Distributions comparison between predicted and ground truth for Test Case 1 (experimental microstructures) at $t = \SI{59}{\minute}$: (a) surface-weighted ECR distributions; (b) grain neighbor count distributions.}
    \label{fig:result_case1_dist}
\end{figure}

Turning to Test Case 1, this scenario addressed a morphologically distinct challenge, where rough and irregular grain boundaries characteristic of experimental microstructures are entirely absent from the synthetically generated training data. The evaluation examined whether this morphological mismatch degraded prediction accuracy and whether the boundary-masked attention mechanism provided any compensatory improvement. Table~\ref{tab:case1_metrics} and the pixel-wise error heatmaps in Figure~\ref{fig:result_case1_heatmap}(a,b) demonstrate that the baseline model achieved $\text{MSE}_{\text{b}}$ of \num{0.3064}, $\text{MAE}_{\text{b}}$ of \num{0.4729}, PSNR of \SI{9.50}{\decibel}, and SSIM of \num{0.5603} at $t = \SI{59}{\minute}$, with residuals broadly distributed across the microstructure. The attention-integrated model achieved lower pixel-wise errors across all metrics, with $\text{MSE}_{\text{b}}$ of \num{0.2722}, $\text{MAE}_{\text{b}}$ of \num{0.4339}, PSNR of \SI{10.70}{\decibel}, and SSIM of \num{0.6593}, indicating that the boundary-masked attention mechanism provided a measurable improvement in pixel-level prediction accuracy for this test case.

At the distributional level, Figure~\ref{fig:result_case1_dist}(a) indicates that the boundary-masked attention model produced substantially lower ECR deviations compared to the baseline, with the $\overline{R}$ error decreasing from \SI{14.40}{\percent} to \SI{7.04}{\percent}, KL divergence from \num{1.5049} to \num{0.0604}, and Wasserstein distance from \num{0.0111} to \num{0.0061}. Furthermore, the grain neighbor count distribution shown in Figure~\ref{fig:result_case1_dist}(b) followed the same pattern, where the attention model recorded KL divergence of \num{0.0313} and Wasserstein distance of \num{0.1509} compared to \num{0.1679} and \num{0.5969} for the baseline. Across all metric categories, the proposed attention mechanism produced lower errors and closer alignment with the ground truth for this test case.

\subsection{Test Case 2: Synthetic Bimodal Microstructures}

\begin{figure}[h!]
    \centering
    \begin{subfigure}[t]{0.48\linewidth}
    \centering
    \includegraphics[width=\linewidth]{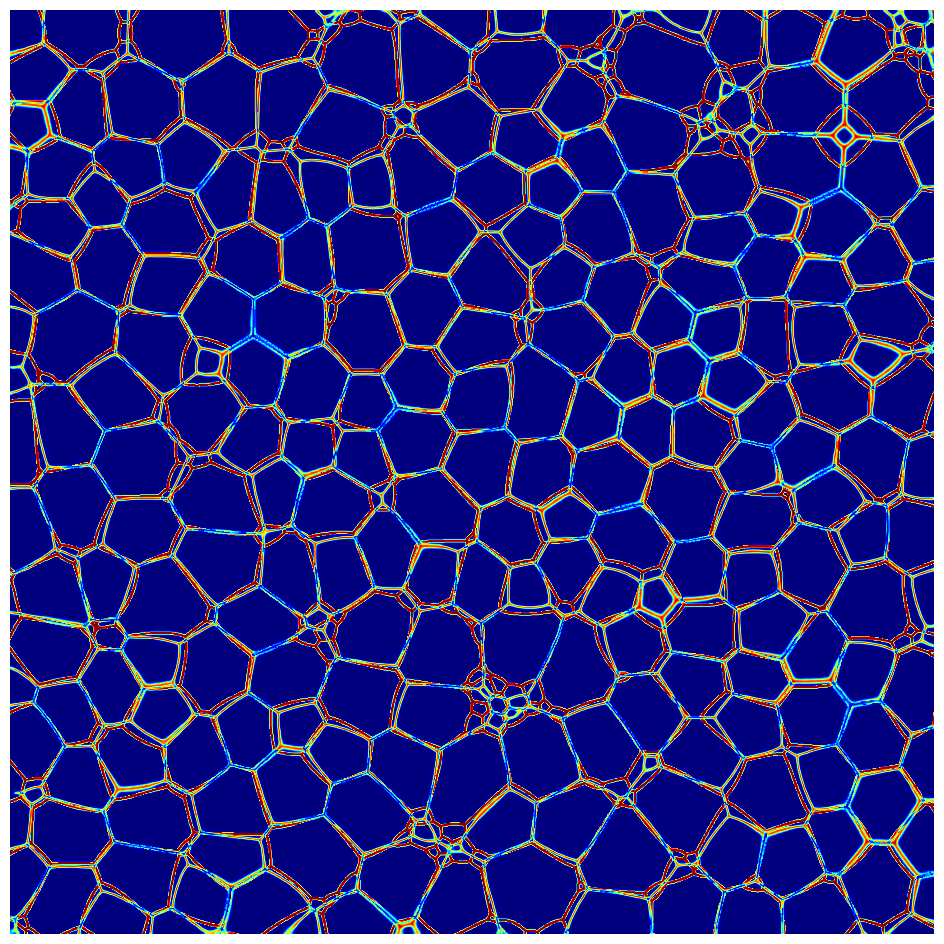}
    \caption{}
    \label{fig:result_case2_heatmap_a}
    \end{subfigure}
    \hfill
    \begin{subfigure}[t]{0.48\linewidth}
    \centering
    \includegraphics[width=\linewidth]{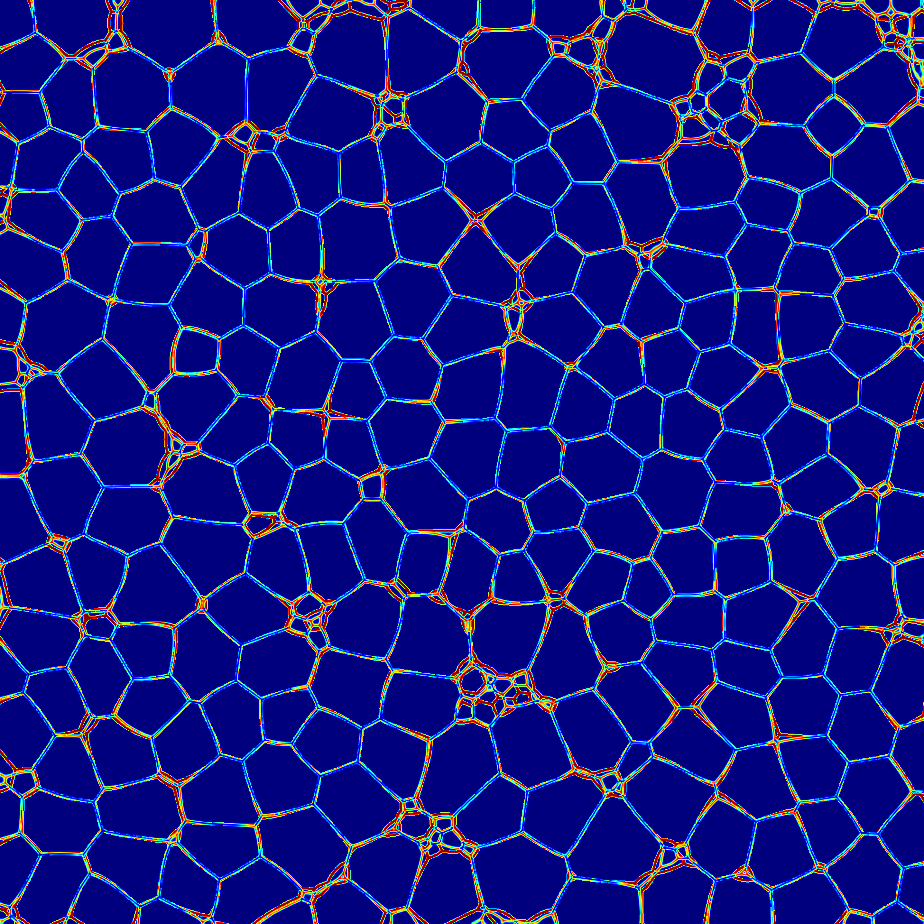}
    \caption{}
    \label{fig:result_case2_heatmap_b}
    \end{subfigure}
    \vspace{0.10cm}
    \centering
    \includegraphics[width=1\textwidth]{figures/heatmap_legend.pdf}
    \caption{Error heatmap for Test Case 2 (synthetic bimodal microstructures) at $t = \SI{59}{\minute}$: (a) baseline model; (b) attention model. (For interpretation of the references to color in this figure legend, the reader is referred to the web version of this article.)}
    \label{fig:result_case2_heatmap}
\end{figure}

\begin{table}[htbp]
    \centering
    \resizebox{\textwidth}{!}{%
    \begin{tabular}{l|cccc|ccc|cc}
    \toprule
    \multirow{2}{*}{\textbf{Model}} & \multicolumn{4}{c|}{\textbf{Pixel-wise and Perceptual}} & \multicolumn{3}{c|}{\textbf{ECR Distribution}} & \multicolumn{2}{c}{\textbf{Neighbor Count Distribution}} \\
    \cmidrule(lr){2-5} \cmidrule(lr){6-8} \cmidrule(lr){9-10}
    & $\text{MSE}_\text{b}$ $\downarrow$ & $\text{MAE}_\text{b}$ $\downarrow$ & PSNR (dB) $\uparrow$ & SSIM $\uparrow$ & $\overline{R}$ Err. (\%) $\downarrow$ & KL $\downarrow$ & W $\downarrow$ & KL $\downarrow$ & W $\downarrow$ \\
    \midrule
    Baseline & 0.3118 & 0.4737 & 10.10 & 0.6221 & 8.75 & 3.1934 & 0.0082 & 0.2296 & 0.3653 \\
    Boundary-Masked Attention & \textbf{0.2171} & \textbf{0.3773} & \textbf{12.451} & \textbf{0.7609} & \textbf{3.57} & \textbf{0.0473} & \textbf{0.0035} & \textbf{0.0799} & \textbf{0.2155} \\
    \bottomrule
    \end{tabular}%
    }

    \vspace{2pt}
    {\footnotesize KL = KL divergence (predicted $\rightarrow$ ground truth); W = Wasserstein distance. Best values are shown in \textbf{bold}.\par}
    \caption{Quantitative evaluation metrics for Test Case 2 (synthetic bimodal microstructures) at $t = \SI{59}{\minute}$.}
    \label{tab:case2_metrics}
\end{table}

\begin{figure}[h!]
    \centering
    \begin{subfigure}[t]{0.48\linewidth}
    \centering
    \includegraphics[width=\linewidth]{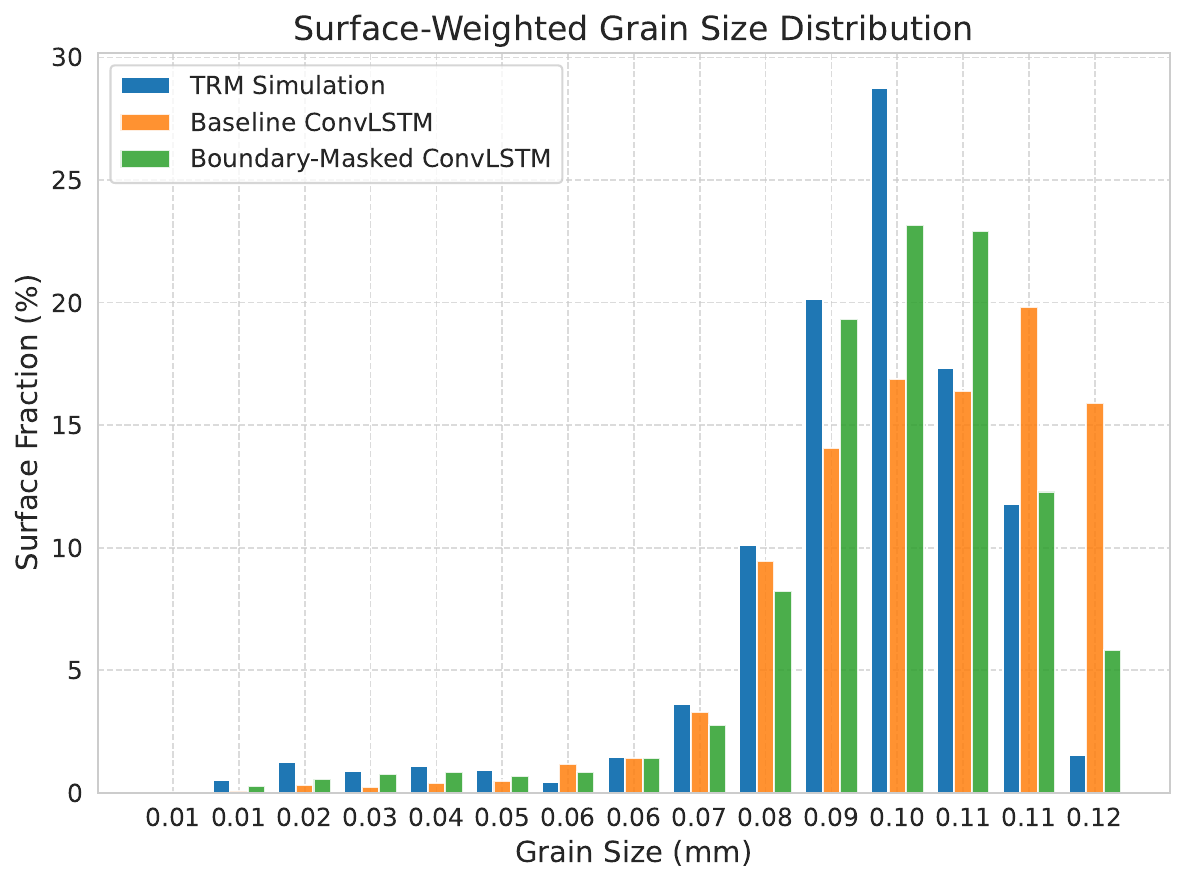}
    \caption{}
    \label{fig:result_case2_dist_a}
    \end{subfigure}
    \hfill
    \begin{subfigure}[t]{0.48\linewidth}
    \centering
    \includegraphics[width=\linewidth]{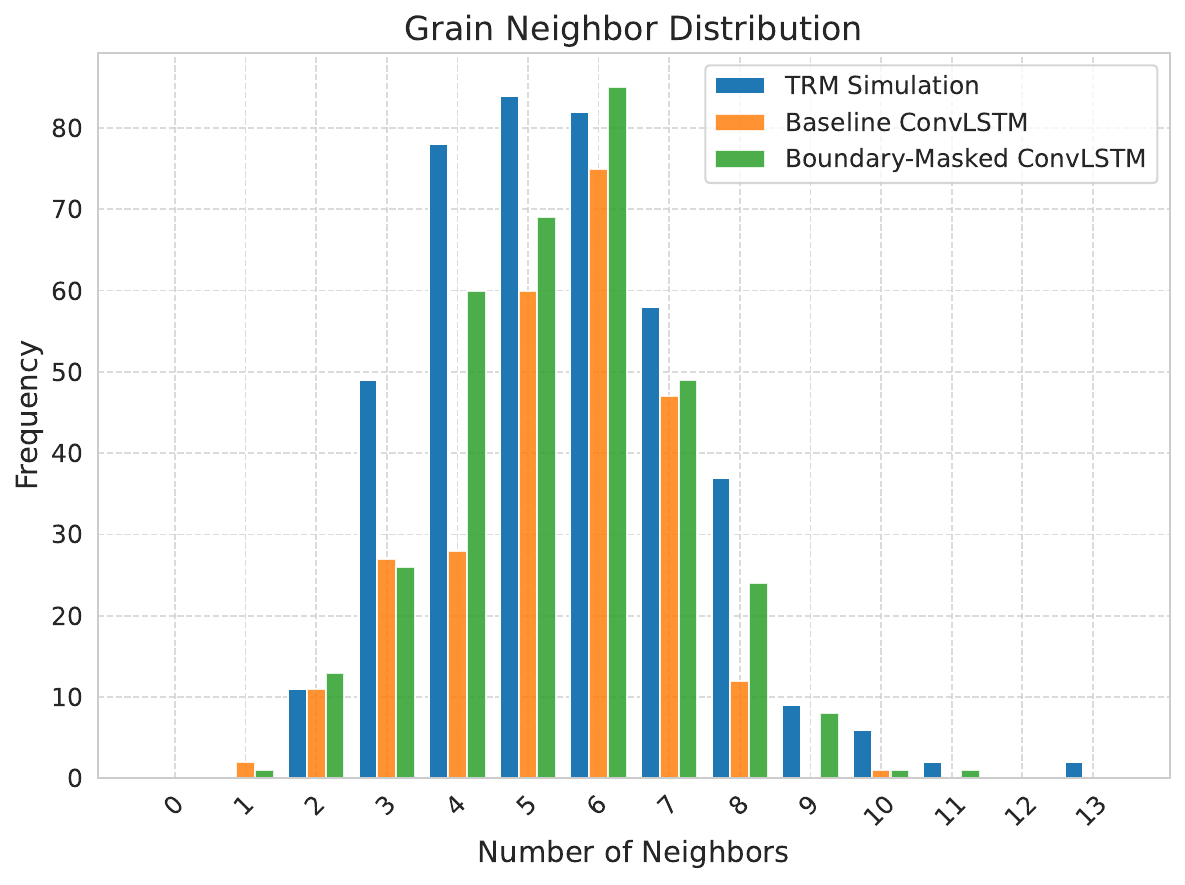}
    \caption{}
    \label{fig:result_case2_dist_b}
    \end{subfigure}
    \caption{Distributions comparison between predicted and ground truth for Test Case 2 (synthetic bimodal microstructures) at $t = \SI{59}{\minute}$: (a) surface-weighted ECR distributions; (b) grain neighbor count distributions.}
    \label{fig:result_case2_dist}
\end{figure}

In contrast to the previous test case, the second test case introduced a statistical rather than morphological challenge, examining model generalization to microstructures characterized by a bimodal initial grain size distribution that differs from the unimodal Gaussian distribution of the training data. This scenario operated within the same simulation framework as the training data, without the added complexity of irregular boundary morphology. As summarized in Table~\ref{tab:case2_metrics}, the attention-integrated model achieved substantially lower pixel-wise errors than the baseline. The $\text{MSE}_{\text{b}}$ decreased from \num{0.3118} to \num{0.2171}, the $\text{MAE}_{\text{b}}$ decreased from \num{0.4737} to \num{0.3773}, PSNR increased from \SI{10.10}{\decibel} to \SI{12.45}{\decibel}, and SSIM increased from \num{0.6221} to \num{0.7609}, as shown in the corresponding error heatmaps in Figure~\ref{fig:result_case2_heatmap}(a,b).

Turning to distributional metrics, Figure~\ref{fig:result_case2_dist}(a) reveals that the boundary-masked attention model closely reproduced the ground truth ECR distribution, achieving $\overline{R}$ error of \SI{3.57}{\percent}, KL divergence of \num{0.0473}, and Wasserstein distance of \num{0.0035}. In comparison, the baseline produced substantially higher deviations of \SI{8.75}{\percent}, \num{3.1934}, and \num{0.0082} for the same metrics. The grain neighbor count distribution in Figure~\ref{fig:result_case2_dist}(b) followed a consistent trend, where the proposed model recorded KL divergence of \num{0.0799} and Wasserstein distance of \num{0.2155} while the baseline recorded \num{0.2296} and \num{0.3653}, respectively. Collectively, these results indicate that the attention mechanism produced better agreement with the ground truth across both pixel-level and distributional metrics for this test case.

\subsection{Test Case 3: Abnormal Grain Growth}

\begin{figure}[h!]
    \centering
    \begin{subfigure}[t]{0.48\linewidth}
    \centering
    \includegraphics[width=\linewidth]{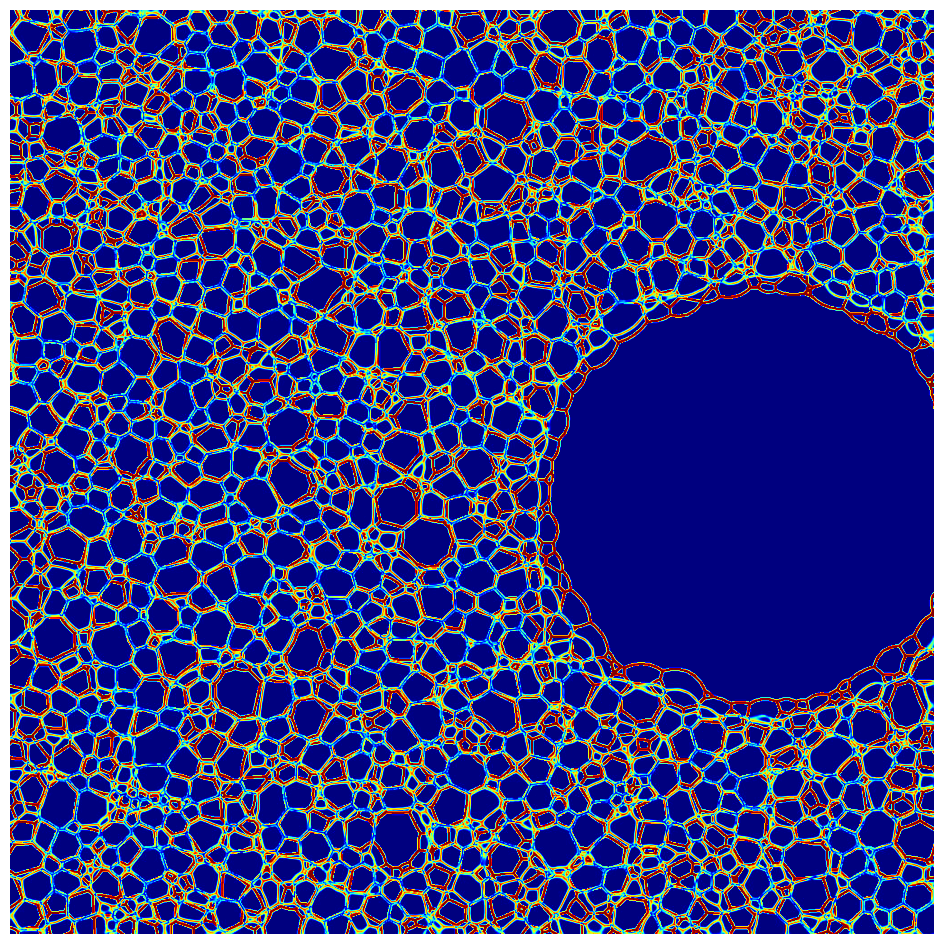}
    \caption{}
    \label{fig:result_case3_heatmap_a}
    \end{subfigure}
    \hfill
    \begin{subfigure}[t]{0.48\linewidth}
    \centering
    \includegraphics[width=\linewidth]{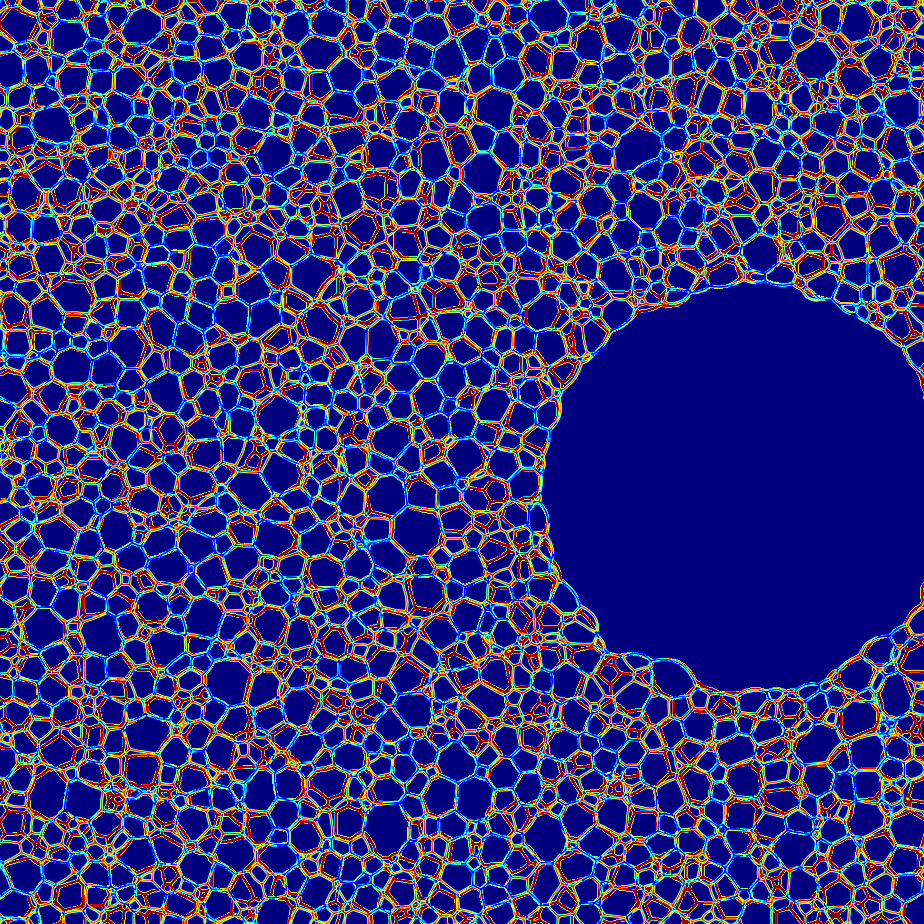}
    \caption{}
    \label{fig:result_case3_heatmap_b}
    \end{subfigure}
    \vspace{0.10cm}
    \centering
    \includegraphics[width=1\textwidth]{figures/heatmap_legend.pdf}
    \caption{Error heatmap for Test Case 3 (abnormal grain growth) at $t = \SI{59}{\minute}$: (a) baseline model; (b) attention model. (For interpretation of the references to color in this figure legend, the reader is referred to the web version of this article.)}
    \label{fig:result_case3_heatmap}
\end{figure}

\begin{table}[htbp]
    \centering
    \resizebox{\textwidth}{!}{%
    \begin{tabular}{l|cccc|ccc|cc}
    \toprule
    \multirow{2}{*}{\textbf{Model}} & \multicolumn{4}{c|}{\textbf{Pixel-wise and Perceptual}} & \multicolumn{3}{c|}{\textbf{ECR Distribution}} & \multicolumn{2}{c}{\textbf{Neighbor Count Distribution}} \\
    \cmidrule(lr){2-5} \cmidrule(lr){6-8} \cmidrule(lr){9-10}
    & $\text{MSE}_\text{b}$ $\downarrow$ & $\text{MAE}_\text{b}$ $\downarrow$ & PSNR (dB) $\uparrow$ & SSIM $\uparrow$ & $\overline{R}$ Err. (\%) $\downarrow$ & KL $\downarrow$ & W $\downarrow$ & KL $\downarrow$ & W $\downarrow$ \\
    \midrule
    Baseline & 0.2720 & 0.4370 & \textbf{7.71} & 0.4016 & 14.97 & 4.5844 & 0.0207 & 0.0294 & 0.1108 \\
    Boundary-Masked Attention & \textbf{0.2703} & \textbf{0.4328} & 7.66 & \textbf{0.4101} & \textbf{2.11} & \textbf{0.0073} & \textbf{0.0064} & \textbf{0.0059} & \textbf{0.0635} \\
    \bottomrule
    \end{tabular}%
    }

    \vspace{2pt}
    {\footnotesize KL = KL divergence (predicted $\rightarrow$ ground truth); W = Wasserstein distance. Best values are shown in \textbf{bold}.\par}
    \caption{Quantitative evaluation metrics for Test Case 3 (abnormal grain growth) at $t = \SI{59}{\minute}$.}
    \label{tab:case3_metrics}
\end{table}

\begin{figure}[h!]
    \centering
    \begin{subfigure}[t]{0.48\linewidth}
    \centering
    \includegraphics[width=\linewidth]{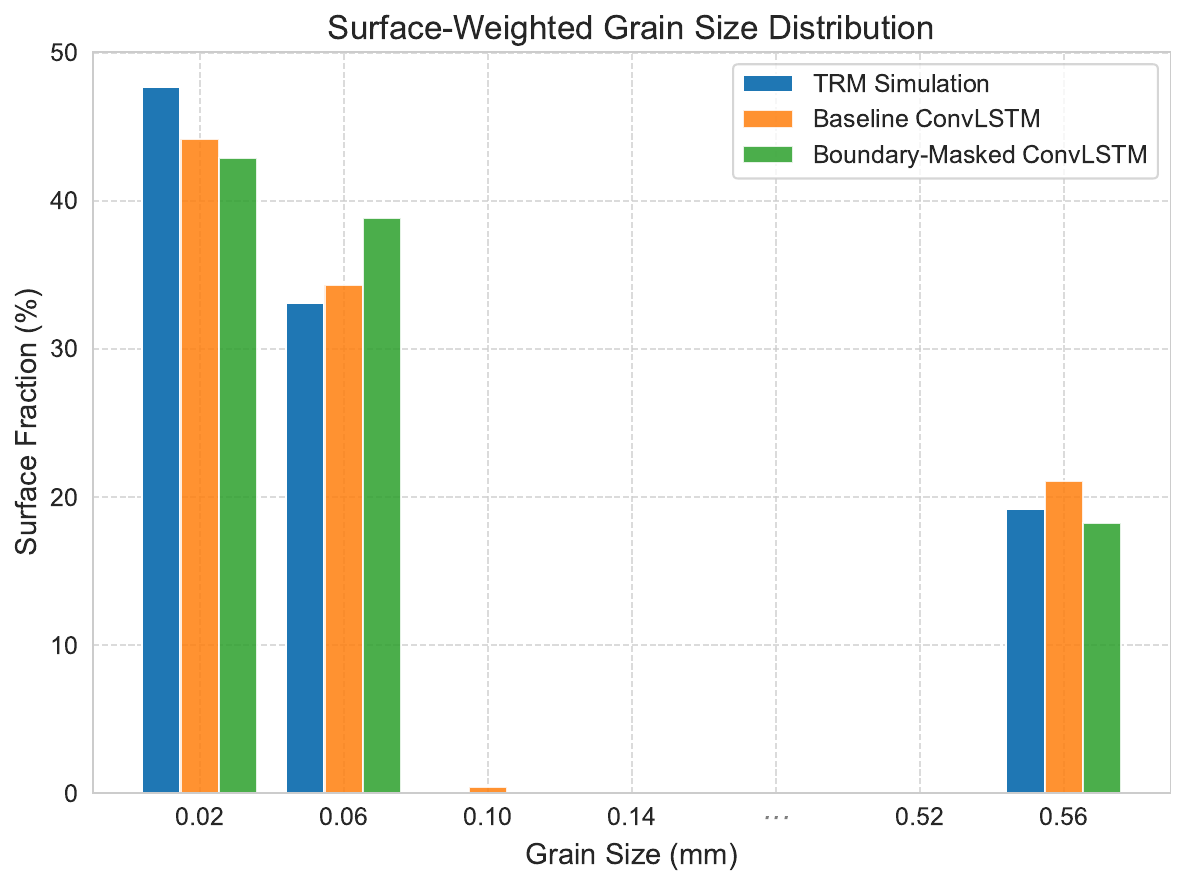}
    \caption{}
    \label{fig:result_case3_dist_a}
    \end{subfigure}
    \hfill
    \begin{subfigure}[t]{0.48\linewidth}
    \centering
    \includegraphics[width=\linewidth]{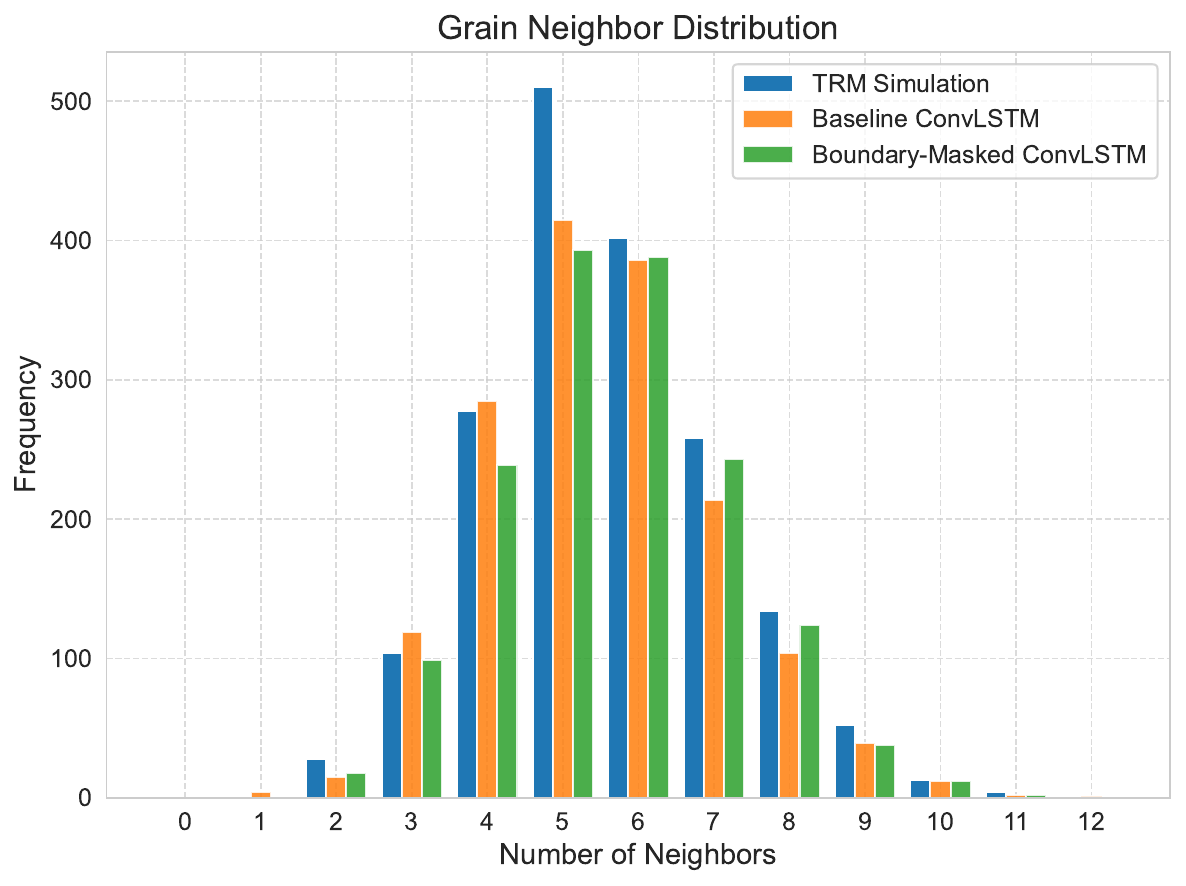}
    \caption{}
    \label{fig:result_case3_dist_b}
    \end{subfigure}
    \caption{Distributions comparison between predicted and ground truth for Test Case 3 (abnormal grain growth) at $t = \SI{59}{\minute}$: (a) surface-weighted ECR distributions; (b) grain neighbor count distributions, with the abnormal grain excluded from the distribution.}
    \label{fig:result_case3_dist}
\end{figure}

Extending the evaluation to the most physically distinct scenario, the third test case represented the most fundamental deviation from the training conditions, examining model generalization to abnormal grain growth, where spatially heterogeneous boundary mobility drives a coarsening mechanism qualitatively different from the average grain growth behavior on which both models were trained. As listed in Table~\ref{tab:case3_metrics}, the attention model showed modest pixel-wise improvements over the baseline in most metrics, with $\text{MSE}_{\text{b}}$ of \num{0.2703} versus \num{0.2720}, $\text{MAE}_{\text{b}}$ of \num{0.4328} versus \num{0.4370}, and SSIM of \num{0.4101} versus \num{0.4016}, while PSNR was marginally lower at \SI{7.66}{\decibel} compared to \SI{7.71}{\decibel} for the baseline, as shown in the corresponding error heatmaps in Figure~\ref{fig:result_case3_heatmap}(a,b).

With respect to distributional accuracy, Figure~\ref{fig:result_case3_dist}(a) reveals a bin-dependent pattern in which the two models differ in the location of their deviations from the ground truth ECR distribution. At the smaller grain size bins, the boundary-masked attention model showed larger deviations from the ground truth than the baseline, indicating that the proposed attention mechanism does not improve the reproduction of the fine grain fraction in this scenario. However, at the large-grain tail bin corresponding to the abnormal grain, the boundary-masked model produced a smaller deviation from the ground truth than the baseline, suggesting that the proposed attention mechanism captures the coarsening of the large abnormal grain more accurately. Despite this bin-level difference in where each model deviates, the aggregate distributional metrics remained lower for the attention model, with the $\overline{R}$ error decreasing from \SI{14.97}{\percent} to \SI{2.11}{\percent}, KL divergence from \num{4.5844} to \num{0.0073}, and Wasserstein distance from \num{0.0207} to \num{0.0064}. In contrast, the grain neighbor count distribution in Figure~\ref{fig:result_case3_dist}(b) showed that the attention-integrated model achieved more consistent agreement with the ground truth across the histogram bins than the baseline, with the attention model producing more accurate predictions for grains with six or more neighbors while the baseline showed closer agreement for grains with five or fewer neighbors, with the proposed model recording KL divergence of \num{0.0059} and Wasserstein distance of \num{0.0635} compared to \num{0.0294} and \num{0.1108} for the baseline.

\section{Discussion}

\subsection{Generalization Capability}

Across all three test cases, the results demonstrated that both models, the baseline and the boundary-masked attention models, generalized to the OOD scenarios without retraining or fine-tuning. This suggests that the neural networks learned the representations and captured the underlying physics of curvature-driven boundary migration rather than patterns specific to the training data. This generalization capability could be attributed to the encoder-decoder architecture (Figure~\ref{fig:architecture}), whose latent representations retained fundamental spatiotemporal features such as boundary curvature and local grain topology that appeared to transfer across different boundary morphologies, grain size distributions, and mobility characteristics.

More specifically, in the bimodal and abnormal grain growth cases, generalization arose due to the fact that local curvature-driven migration remains the governing mechanism regardless of the global distribution shape or mobility heterogeneity. The experimental microstructure case followed a different physical logic, where isothermal annealing progressively smoothed the initially rough grain boundaries, driving the morphology toward the smooth configuration characteristic of the synthetic training data. Prediction accuracy therefore improved over the horizon, with early steps presenting the most challenging conditions and later steps benefiting from the growing alignment between the evolving morphology and the training domain.

\subsection{Effect of Attention Mechanism on Performance}
\label{sec:attention_effect}

\begin{figure}[h!]
    \centering
    \resizebox{0.70\textwidth}{!}{%
    \begin{minipage}{1.0\textwidth}
    \centering
    \begin{subfigure}[b]{0.49\textwidth}
        \centering
        \includegraphics[width=\textwidth]{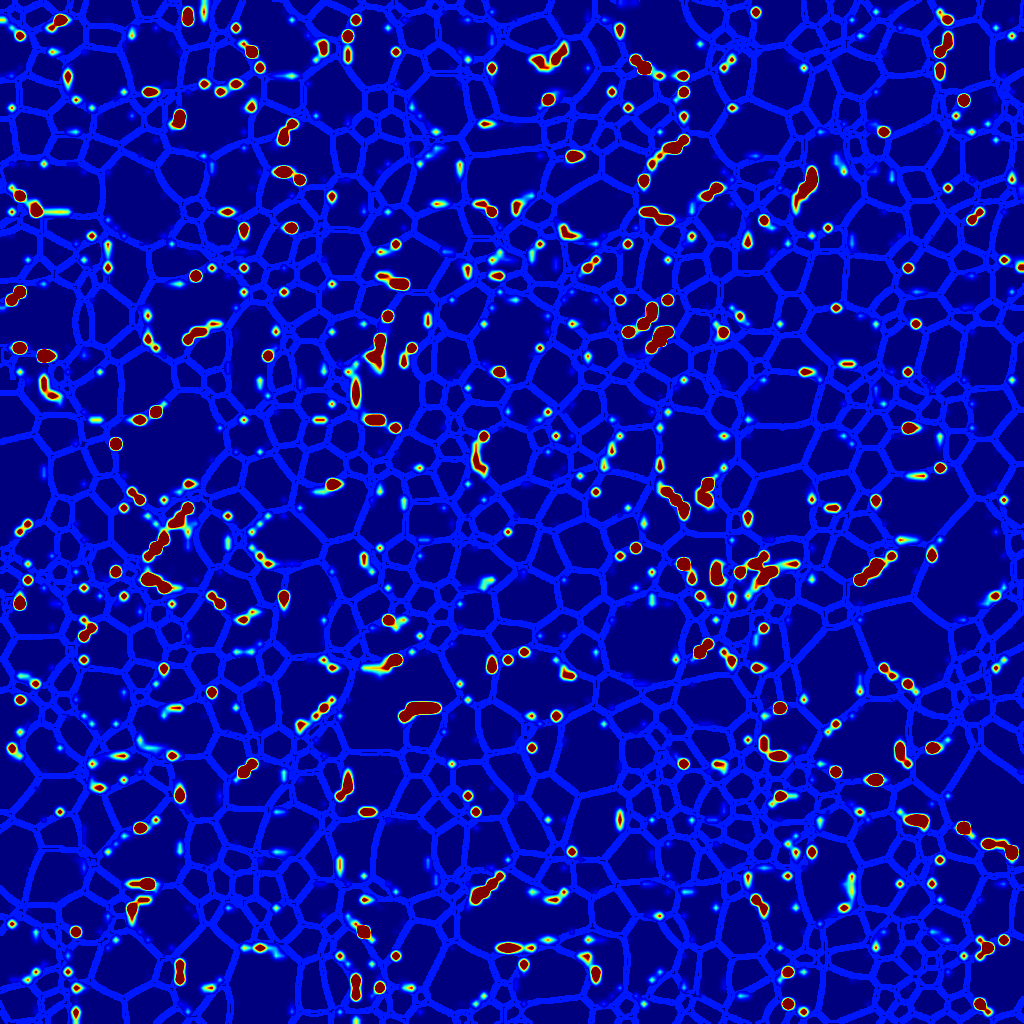}
        \caption{}
    \end{subfigure}
    \hfill
    \begin{subfigure}[b]{0.49\textwidth}
        \centering
        \includegraphics[width=\textwidth]{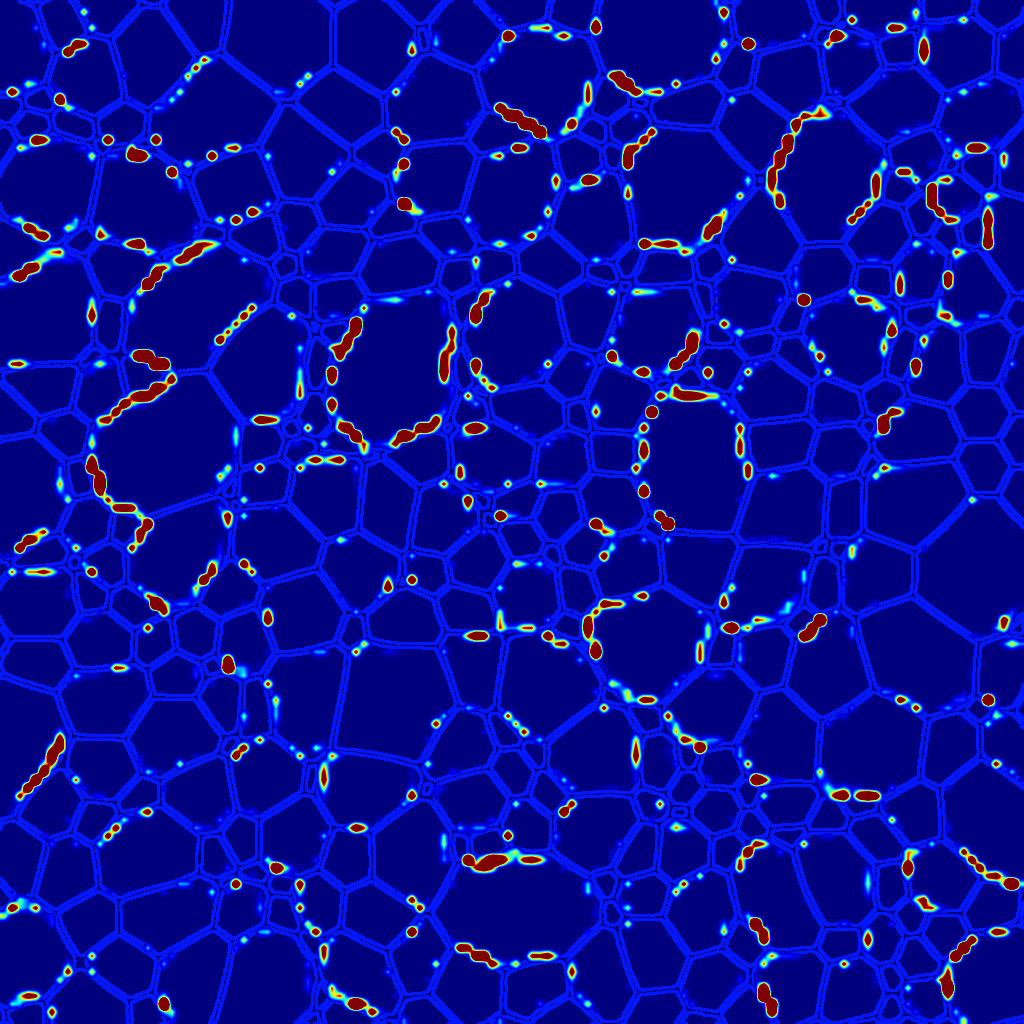}
        \caption{}
    \end{subfigure}

    \vspace{0.5em}

    \begin{subfigure}[b]{0.49\textwidth}
        \centering
        \includegraphics[width=\textwidth]{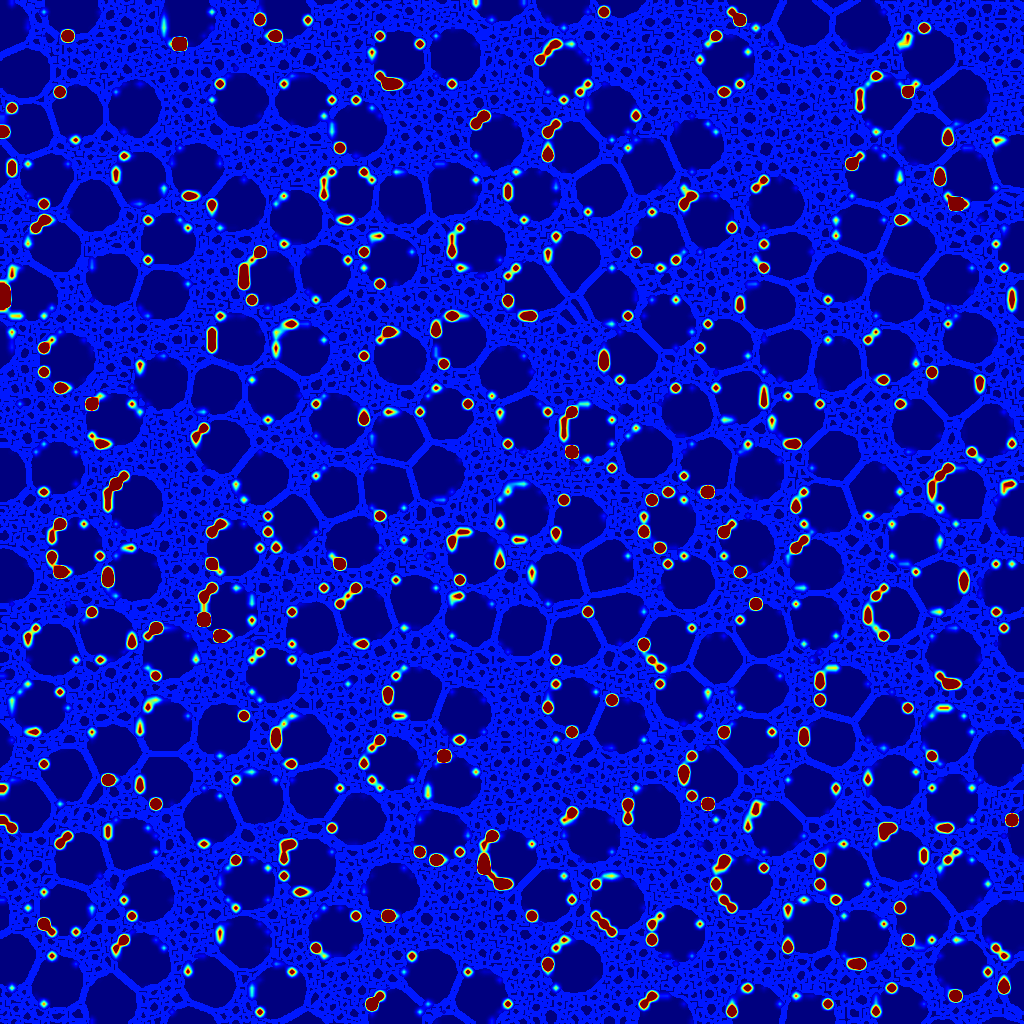}
        \caption{}
    \end{subfigure}
    \hfill
    \begin{subfigure}[b]{0.49\textwidth}
        \centering
        \includegraphics[width=\textwidth]{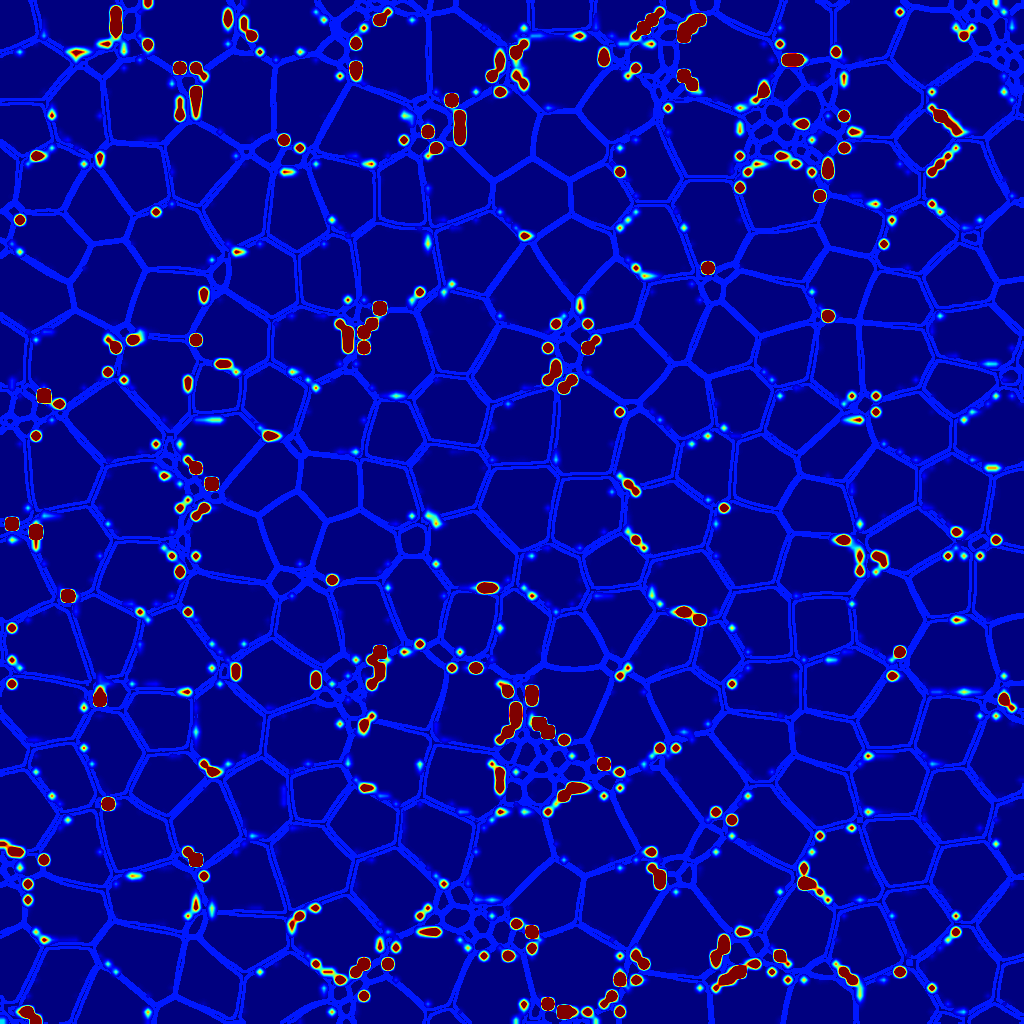}
        \caption{}
    \end{subfigure}

    \vspace{0.5em}

    \begin{subfigure}[b]{0.49\textwidth}
        \centering
        \includegraphics[width=\textwidth]{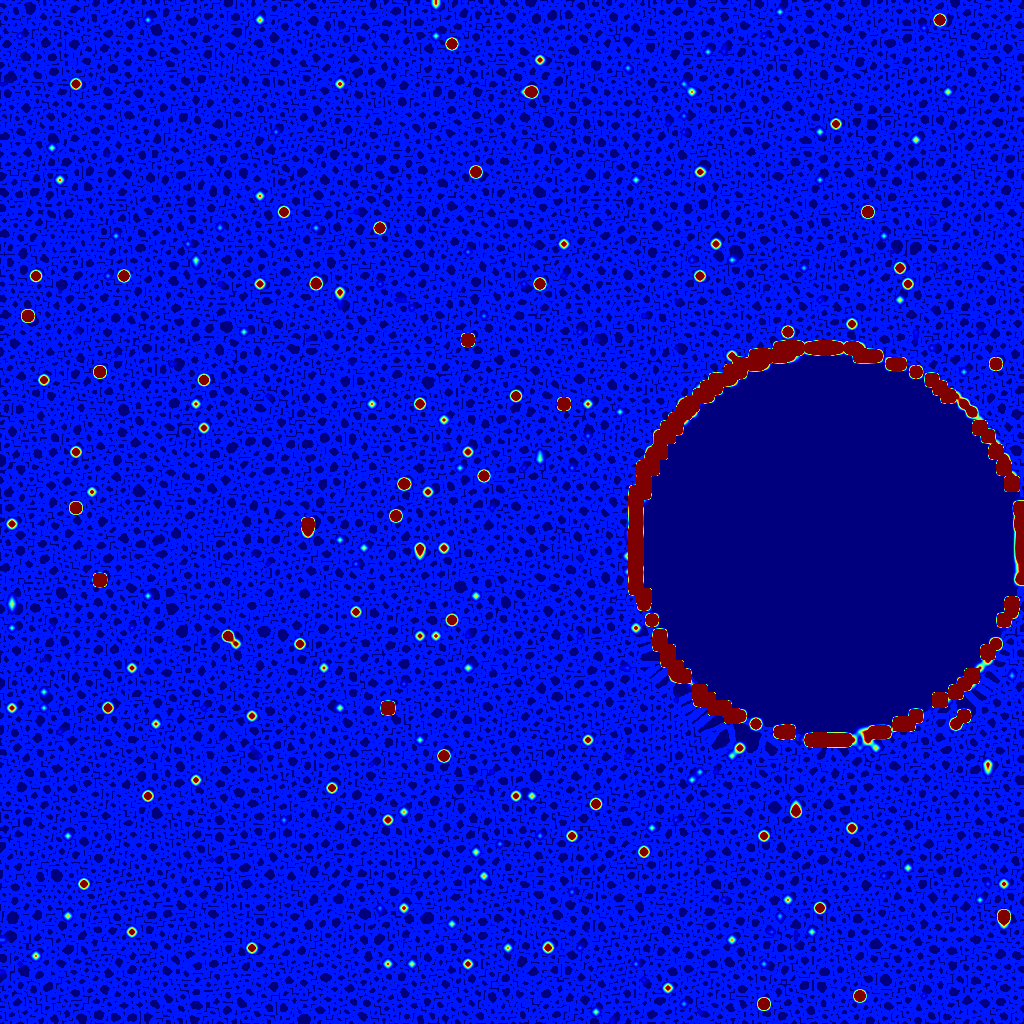}
        \caption{}
    \end{subfigure}
    \hfill
    \begin{subfigure}[b]{0.49\textwidth}
        \centering
        \includegraphics[width=\textwidth]{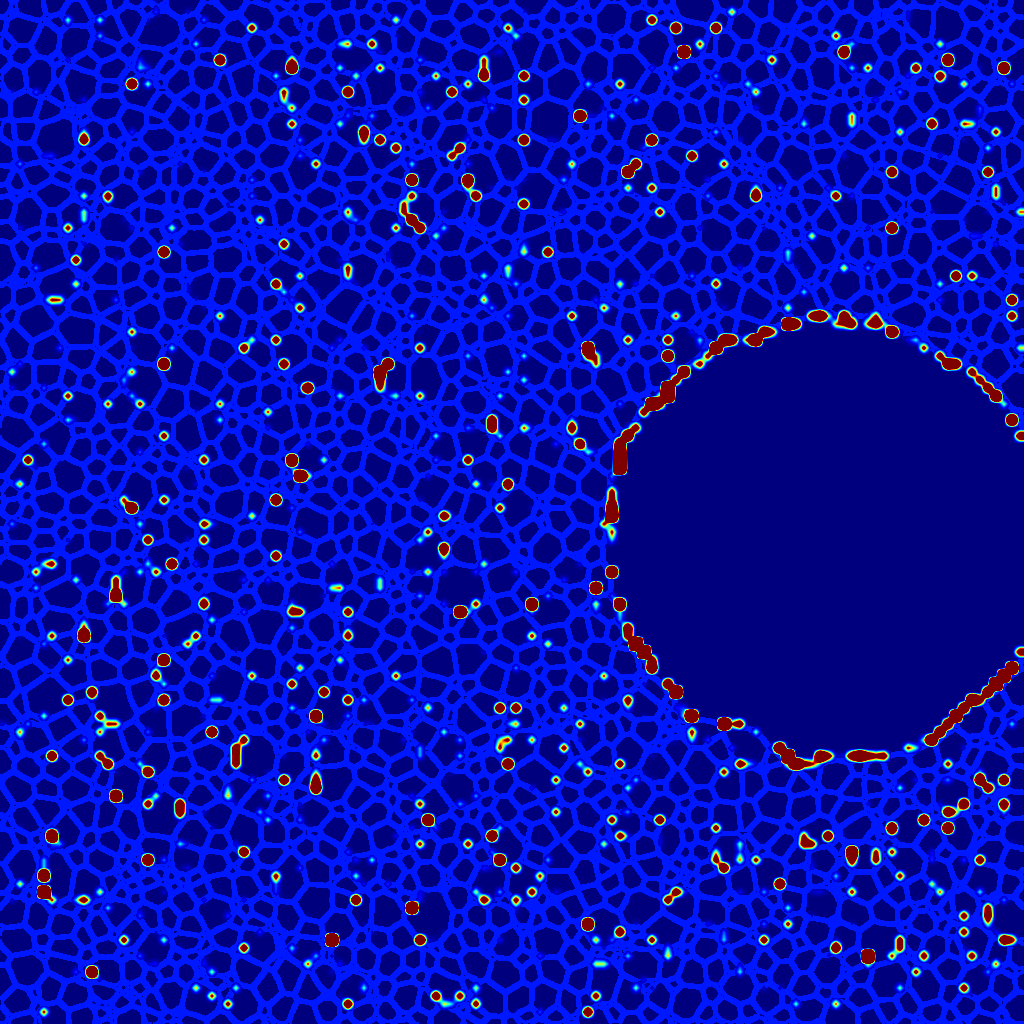}
        \caption{}
    \end{subfigure}
    \end{minipage}%
    }

    \vspace{0.10cm}

    \centering
    \includegraphics[width=1\textwidth]{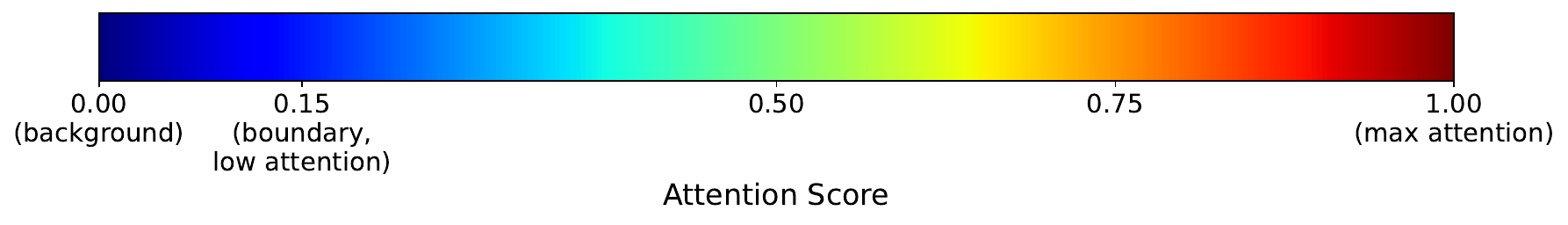}

    \caption{Attention weight heatmaps generated by the boundary-masked attention mechanism at the first prediction step ($t = \SI{10}{\minute}$) and the final prediction step ($t = \SI{59}{\minute}$): (a, b) Test Case 1 (experimental microstructures); (c, d) Test Case 2 (synthetic bimodal microstructures); (e, f) Test Case 3 (abnormal grain growth). Brighter regions indicate higher attention weights, with the corresponding legend shown below. (For interpretation of the references to color in this figure legend, the reader is referred to the web version of this article.)}
    \label{fig:attention_heatmaps_all_cases}
\end{figure}

The proposed boundary-masked attention mechanism provided consistent improvement over the baseline across all three test cases, while also offering interpretable insight into where the model directed its attention during prediction. Improvement in pixel-wise metrics was more pronounced for the second test case than for the others, where large smooth grain boundaries dominated the coarsening dynamics from the earliest prediction steps, while distributional accuracy improved across all cases. As shown in Figure~\ref{fig:attention_heatmaps_all_cases}, the proposed mechanism consistently allocated higher attention weights to the boundaries of larger grains across all test cases. This size-dependent weighting was not explicitly designed but emerged from training, as large grain boundaries persisted throughout the evolution and carried more predictive signal for the long-term coarsening trajectory than the boundaries of subcritical grains that disappear early in the coarsening sequence.

Regarding the experimental microstructure, rough boundaries at early prediction steps produced spatially dispersed and less coherent attention weights (Figure~\ref{fig:attention_heatmaps_all_cases}(a)). At later steps (Figure~\ref{fig:attention_heatmaps_all_cases}(b)), as boundaries converged toward smoother configurations, attention showed clearer concentration on large grain boundaries. However, in autoregressive prediction, errors from early misaligned attention steps propagated and accumulated through subsequent predictions, leading to increased prediction errors through subsequent steps. Nevertheless, the attention model achieved lower errors than the baseline across all metric categories at the final state of evolution.

In line with this, the bimodal microstructure case (Figure~\ref{fig:attention_heatmaps_all_cases}(c,d)) showed that higher attention weights were concentrated along the boundaries of the large grains from the first prediction step, while the small grain population received substantially lower weights. This size-dependent weighting pattern is consistent with the large-grain boundary dominance described above, where lower boundary curvature corresponds to more persistent configurations that carry more predictive signal over extended horizons. At the final prediction step, this concentration was maintained and progressively intensified as small grain boundaries disappeared from the boundary mask, enabling accurate tracking of the coarsening trajectory throughout the prediction horizon.

For the case of abnormal grain growth (Figure~\ref{fig:attention_heatmaps_all_cases}(e,f)), the attention mechanism similarly concentrated on the boundary of the abnormal grain as the largest and most persistent feature in the microstructure, consistent with the size-dependent weighting pattern described above. However, unlike the bimodal microstructure case where small grains disappeared progressively from the boundary mask, the major normal grains surrounding the abnormal grain in this case were also actively evolving through continuous growth and topological changes throughout the prediction horizon. By learning to concentrate attention on the large grain boundary of the abnormal grain, the model allocated comparatively lower weights to these actively changing normal grain boundaries, which accounts for the modest pixel-wise improvements observed for this test case. 

\subsection{ECR and Neighbor Count Distribution}

Complementing the pixel-level metrics, the surface-weighted ECR distribution and grain neighbor count distribution analysis further characterize the quality of the predicted microstructures. The distributions were extracted through an image processing pipeline of microstructure image binarization and grain segmentation, inheriting the over-segmentation and under-segmentation limitations documented in the previous study. Over-segmentation artificially inflates the detected grain count and underestimates grain sizes, while under-segmentation reduces the detected grain count and overestimates grain sizes, introducing discrepancies in the extracted distributions that are independent of boundary prediction quality.

Starting with the experimental microstructure case, both models captured the general ECR distribution shape, with the attention model showing closer agreement across grain sizes from \SIrange{0.01}{0.15}{\mm}, where the baseline model produced substantially larger deviations. However, both models overpredicted the frequency of grains above approximately \SI{0.17}{\mm}, as the boundary topology at these sizes was not accurately predicted. For the grain neighbor count distribution in Figure~\ref{fig:result_case1_dist}(b), the attention model maintained closer agreement with the ground truth across most bins, while the baseline model underpredicted the frequency of grains with neighbor counts of \numrange{5}{8}.

A similar trend emerged for the bimodal microstructure case, where the attention model showed closer agreement with the ground truth ECR distribution across grain sizes from \SIrange{0.01}{0.09}{\mm} and above \SI{0.11}{\mm}, while the baseline model underpredicted grain frequency at sizes of \SIrange{0.09}{0.11}{\mm}. However, both models overpredicted grain frequency at approximately \SI{0.12}{\mm}, with the attention model producing a smaller deviation than the baseline, consistent with the experimental microstructure case. This persistent overprediction of large grain frequency, despite the attention mechanism allocating higher weights to large grain boundaries, is attributed to the fact that large grains share many boundary junctions with the surrounding small grains. At these junctions, lower attention weights on small grain boundaries lead to less accurate local predictions, resulting in residual errors in grain boundary closure during segmentation. At the topological level, the attention model showed closer agreement with the ground truth across all neighbor count bins with the exception of neighbor count \num{3}, while the baseline model underpredicted the frequency of grains with neighbor counts of \numrange{4}{9}, indicating improved accuracy in capturing the topological evolution of the microstructure.

\begin{figure}[h!]
    \centering
    \includegraphics[width=\textwidth]{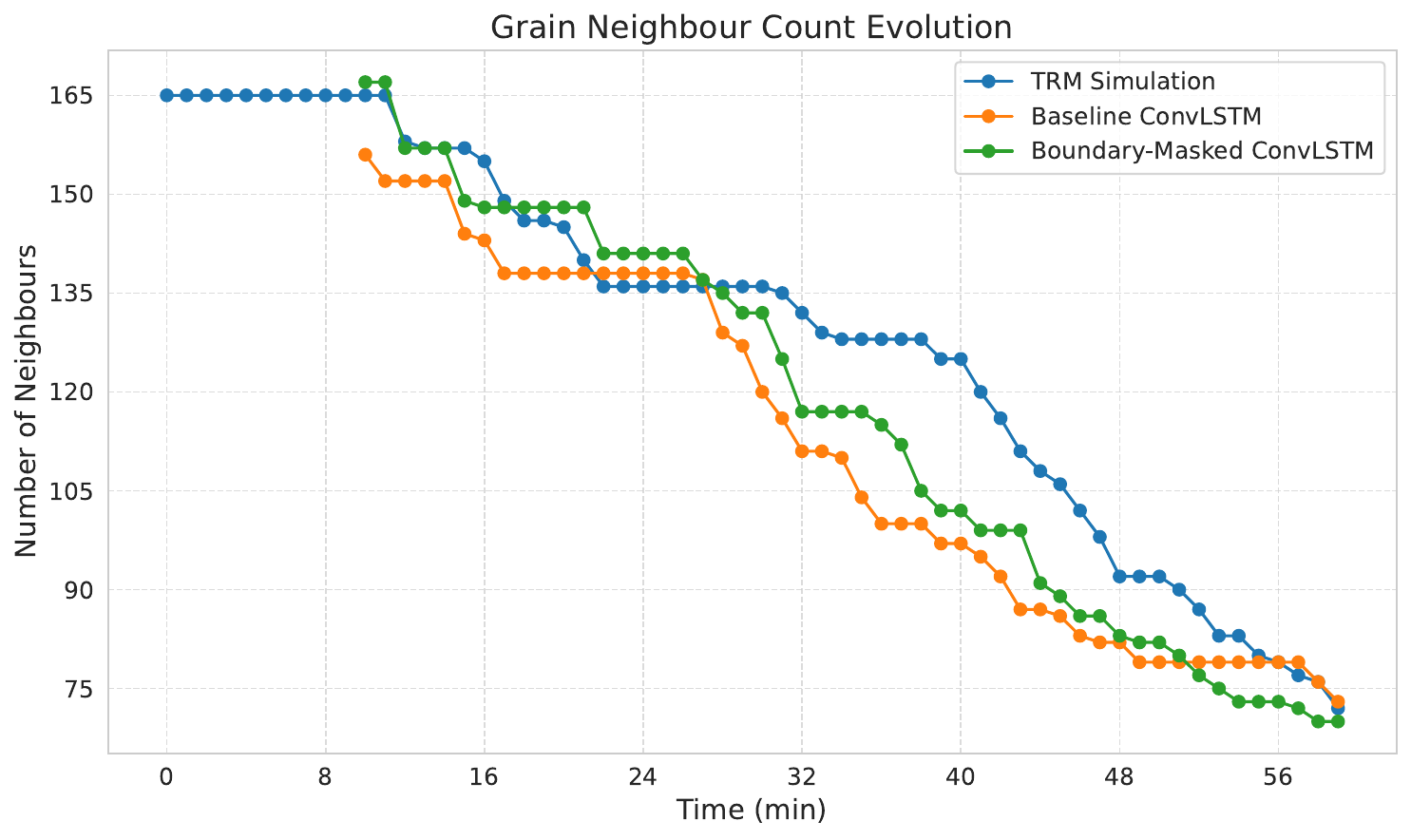}
    \caption{Evolution of the neighbor count for the single abnormal grain in Test Case 3 from $t = \SI{0}{\minute}$ to $t = \SI{59}{\minute}$, defined as the number of normal grains sharing a boundary with the abnormal grain at each timestep, comparing predictions from the baseline and attention models against the ground truth TRM simulation.}
    \label{fig:neighbour_count_evolution}
\end{figure}

Lastly, the abnormal grain growth case revealed a divergent pattern in the ECR distribution, where the boundary-masked attention model produced accurate predictions only at the grain size corresponding to the abnormal grain at approximately \SI{0.56}{\mm}. The baseline model, in contrast, showed closer agreement for the smaller normal grain population in the range of \SIrange{0.02}{0.06}{\mm}. This outcome is consistent with the attention heatmaps and the modest pixel-wise improvements discussed previously, due to the fact that the model learned to concentrate on the large abnormal grain boundary while allocating lower weights to the surrounding normal grains. Despite this, the grain neighbor count distribution in Figure~\ref{fig:result_case3_dist}(b), constructed excluding the abnormal grain, showed that the boundary-masked attention model achieved better accuracy than the baseline across all neighbor count bins. This is attributed to the neighbor count statistics of the normal grain population being predominantly governed by the topological evolution driven by the growth of the abnormal grain, where accurate prediction of the abnormal grain boundary translates into more accurate prediction of grain consumption events and the resulting changes in neighbor connectivity among the surrounding normal grains, as further illustrated by the neighbor count evolution in Figure~\ref{fig:neighbour_count_evolution}.

\subsection{Training and Inference Time}

Regarding training time, the introduced boundary-masked attention model required approximately \num{5} hours compared to \num{3} hours for the baseline. This increase arose as attention weight computation, including score calculations and weighted context summation, was performed at each timestep throughout the processing sequence, accumulating additional matrix operations over the full training dataset. Since training is a one-time cost, this was not a substantial increase relative to the baseline training time. In terms of inference time, both models remained on the order of seconds. The attention model took approximately \SI{15}{\second} to complete the full \SI{50}{\minute} autoregressive prediction compared to approximately \SI{10}{\second} for the baseline, both providing substantial computational advantages over the TRM simulation.

\subsection{Limitations and Outlook}

Despite these findings, the primary limitation of the proposed boundary-masked attention mechanism lies in its inherent bias toward large grain boundaries, which emerged from training due to the fact that large boundaries have lower curvature and therefore migrate more slowly by the capillarity-driven relation at high temperature. This slower migration causes large boundaries to persist throughout the prediction horizon and carry more predictive signal for long-range coarsening trajectories than small grain boundaries that disappear early in the evolution. While this size-dependent weighting benefited overall prediction accuracy, it may underweight the regions where the most active evolution of microstructures occurs at early prediction steps, particularly in microstructures containing a large number of small grains. In microstructures characterized by a bimodal initial grain size distribution, small grains undergo rapid shrinkage and disappear during the early stages of coarsening, yet these boundaries receive comparatively lower attention weights due to their small size. Similarly, in the abnormal grain growth case, the normal grain population surrounding the abnormal grain undergoes the most topologically significant changes at early steps, while the attention concentrates on the boundary of the large abnormal grain. Future work could explore alternative masking strategies that redirect attention toward these dynamically active regions, such as junction-masked attention focusing on junction points where grain boundary migration is most active, or small-grain-masked attention that selectively weights boundaries of subcritical grains undergoing shrinkage during the early prediction stages. An alternative strategy could incorporate the local grain neighbor count as an additional attention feature, since deviations from the equilibrium six-neighbor configuration correspond to locally higher grain boundary migration kinetics, providing a more physically informed attention signal.

Autoregressive error accumulation also remained a limitation shared with the previous study and with sequence-to-sequence prediction tasks more generally. An alternative approach could involve a hybrid surrogate strategy, where the PDE-based simulation is invoked at selected timesteps within the prediction horizon to correct accumulated prediction errors before returning to the ML model, preventing runaway error accumulation during long autoregressive sequences. Broader generalization to physically more diverse systems, including anisotropic boundary energy, grain growth with second-phase particles, or recrystallization, would likely require additional training data or targeted fine-tuning, as these mechanisms introduce dynamics absent from the current training data.

\section{Conclusions}

In summary, both the baseline and the proposed boundary-masked attention models generalized to all three OOD scenarios without retraining or fine-tuning, confirming that the representations learned from synthetically generated pure grain growth data remained applicable across varied boundary morphologies, grain size distributions, and physical mechanisms. This generalization capability is attributed to the encoder-decoder architecture, which learned spatiotemporal representations of boundary curvature and local grain topology rather than memorizing patterns specific to the training data, enabling transfer across the diverse conditions evaluated in this study.

Building on this, the proposed boundary-masked attention mechanism further improved prediction performance across all three test cases, with the highest pixel-wise improvement observed for the bimodal microstructure case where large smooth grain boundaries dominated the coarsening dynamics from the earliest prediction steps. The mechanism learned to allocate higher attention weights to the boundaries of larger grains, a behavior that emerged from training given that large boundaries have lower curvature, migrate more slowly by the capillarity-driven relation, and therefore contribute more sustained gradient signal throughout the prediction horizon. Despite this improvement, the ECR distribution showed residual deviation at large grain sizes, attributed to the fact that large grains share many boundary junctions with surrounding small grains, where lower attention weights on small grain boundaries introduce errors in grain boundary closure during segmentation, further accumulated by the over-segmentation and under-segmentation limitations of the image processing pipeline documented in the previous study.

However, the inherent bias of the boundary-masked attention mechanism toward large grain boundaries, while beneficial overall, represents a limitation in microstructures where small grains undergo active evolution at early prediction steps. This was illustrated by the modest pixel-wise improvements in the test case of abnormal grain growth, where the normal grain population surrounding the abnormal grain was also actively coarsening throughout the prediction horizon, receiving comparatively lower attention weights. Alternative masking strategies, such as junction-masked attention focusing on boundary junction points where grain boundary migration is most active, or small-grain-masked attention that selectively weights the boundaries of subcritical grains undergoing shrinkage, could provide more targeted attention allocation for such cases and represent a promising direction for future work. More broadly, these findings demonstrate that physics-informed attention mechanisms can improve the generalization of DL models trained on synthetic data to practical microstructure conditions, broadening the applicability of data-driven approaches for industrial materials processing.

\section*{Acknowledgments}
The authors thank ArcelorMittal, Aperam, Aubert \&
Duval, CEA, Constellium, Framatome, and Safran companies and the ANR for their financial support through the DIGIMU consortium and RealIMotion ANR Industrial Chair (Grant No. ANR-22-CHIN-0003). The authors also thank Baptiste Flipon for providing the experimental microstructure data used as the out-of-distribution test case in this study.

\section*{Data Availability}
Data will be made available on request.

\section*{Conflicts of Interest}
The authors declare that they have no known competing financial interests or personal relationships that could have appeared to influence the work reported in this paper.

\bibliographystyle{elsarticle-num}
\bibliography{references}

\end{document}